\RequirePackage{fix-cm}
\pdfoutput=1
\documentclass[natbib,twocolumn]{svjour3}          
\smartqed  
\usepackage{graphicx}
\usepackage{float}
\usepackage{microtype}
\usepackage{xspace}
\usepackage[linesnumbered,ruled,vlined]{algorithm2e}
\usepackage{algpseudocode}

\usepackage{booktabs}
\usepackage{tabularx}
\newcolumntype{Y}{>{\centering\arraybackslash}X}
\usepackage{xspace}
\usepackage{esvect}
\usepackage{xcolor}
\usepackage[pagebackref=true,breaklinks=true,citecolor=tealblue,colorlinks,linkcolor=ruby,bookmarks=false]{hyperref}
\definecolor{ruby}{rgb}{0.88, 0.07, 0.37}
\definecolor{tealblue}{rgb}{0.18, 0.40, 0.46}
\usepackage{array}
\usepackage{multirow}
\newcolumntype{H}{>{\setbox0=\hbox\bgroup}c<{\egroup}@{}}
\usepackage{amsmath}
\usepackage{bm}
\usepackage{bbm}
\usepackage{pythonhighlight}
\usepackage{rotating}
\graphicspath{{./}}
\usepackage{mathrsfs}
\usepackage{hyperref}
\usepackage{bbding}
\usepackage{url}
\usepackage{amsfonts}
\usepackage{amssymb}
\usepackage{colortbl}
\usepackage{color}
\usepackage{fdsymbol}
\newcommand{\Perp}{\perp \!\!\! \perp}
\allowdisplaybreaks[4]

\definecolor{accolor}{RGB}{0,0,255}
\definecolor{grayblue}{RGB}{222, 228, 238}
\definecolor{myred}{RGB}{219, 33, 38}
\definecolor{myblue}{RGB}{31, 45, 210}

\begin{document}

\title{Continual Test-Time Adaptation for Single Image Defocus Deblurring via Causal Siamese Networks}


\author{Shuang~Cui    \and
        Yi~Li         \and
        Jiangmeng~Li  \and
        Xiongxin~Tang \and  
        Bing~Su       \and
        Fanjiang~Xu   \and
        Hui~Xiong
}


\institute{Shuang Cui, Yi Li, Jiangmeng Li, Xiongxin Tang and Fanjiang Xu \at
              National Key Laboratory of Space Integrated Information System, Institute of Software Chinese Academy of Sciences;\\
              University of Chinese Academy of Sciences\\
              \email{cuishuang21@mails.ucas.ac.cn, liyitunan@gmail.com, xiongxin@iscas.ac.cn, fanjiang@iscas.ac.cn, jiangmeng2019@ iscas.ac.cn}          
           \and
        Bing Su \at
              Beijing Key Laboratory of Big Data Management and Analysis Methods, Gaoling School of Artificial Intelligence, Renmin University of China\\
              \email{subingats@gmail.com}
              \and
        Hui Xiong \at
              Thrust of Artificial Intelligence, the Hong Kong University of Science and Technology (Guangzhou);\\
              Department of Computer Science \& Engineering, the Hong Kong University of Science and Technology\\
              \email{xionghui@ust.hk}
              \and
        Shuang Cui, Yi Li and Jiangmeng Li have contributed equally to this work. Corresponding author: Jiangmeng Li.
}

\date{Received: date / Accepted: date}

\def\ourconv{RIConv++\xspace}
\def\smallgap{\vspace{0.05in}}
\maketitle

\begin{abstract}
Single image defocus deblurring (SIDD) aims to restore an all-in-focus image from a defocused one. Distribution shifts in defocused images generally lead to performance degradation of existing methods during out-of-distribution inferences. In this work, we gauge the intrinsic reason behind the performance degradation, which is identified as the heterogeneity of lens-specific point spread functions. Empirical evidence supports this finding, motivating us to employ a continual test-time adaptation (CTTA) paradigm for SIDD. However, traditional CTTA methods, which primarily rely on entropy minimization, cannot sufficiently explore task-dependent information for pixel-level regression tasks like SIDD. To address this issue, we propose a novel Siamese networks-based continual test-time adaptation framework, which adapts source models to continuously changing target domains only requiring unlabeled target data in an online manner. To further mitigate semantically erroneous textures introduced by source SIDD models under severe degradation, we revisit the learning paradigm through a structural causal model and propose \textit{Causal Siamese networks} (CauSiam). Our method leverages large-scale pre-trained vision-language models to derive discriminative universal semantic priors and integrates these priors into Siamese networks, ensuring causal identifiability between blurry inputs and restored images. Extensive experiments demonstrate that CauSiam effectively improves the generalization performance of existing SIDD methods in continuously changing domains.

\end{abstract}

\section{Introduction} 
\label{section:introduction}
A large aperture can increase the amount of light entering the lens, thereby reducing exposure time. However, this also results in a shallower depth of field, which means that only objects in the focal plane will be sharply captured. Objects located far from the focal plane may encounter the circle of confusion \citep{potmesil1981lens}, leading to out-of-focus blur. Single image defocus deblurring (SIDD) restores all-in-focus images from defocused ones, caused by the depth of field. SIDD has wide-ranging applications in high-level vision tasks, including object detection \citep{lin2024universal,oza2023unsupervised}, image classification \citep{wang2017residual}, text recognition \citep{shi2016robust}, and autonomous driving \citep{wang2023multi}.

\begin{figure*}
\centering
\includegraphics[width=0.199\linewidth]{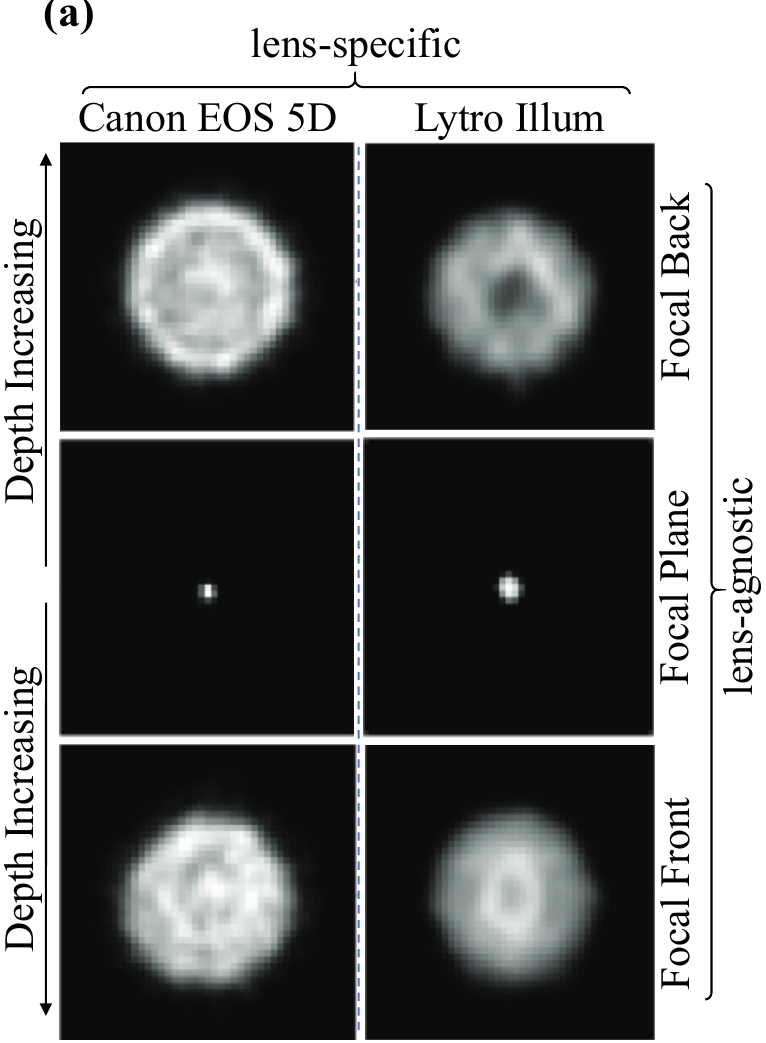}
\includegraphics[width=0.237\linewidth]{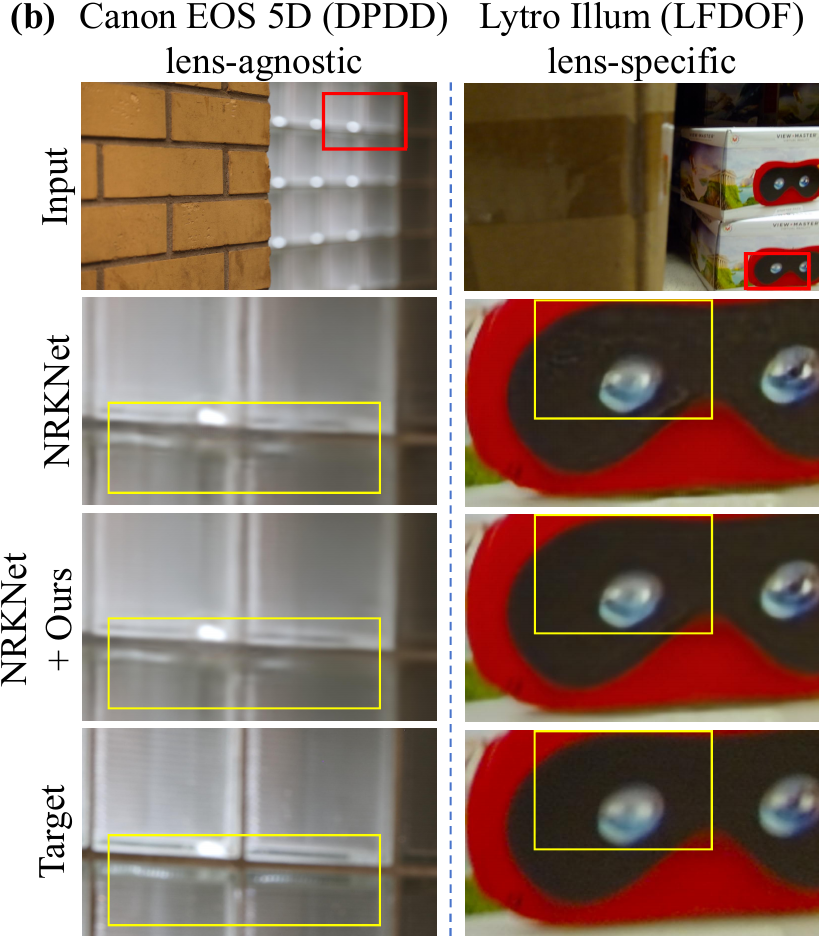}
\includegraphics[width=0.307\textwidth]{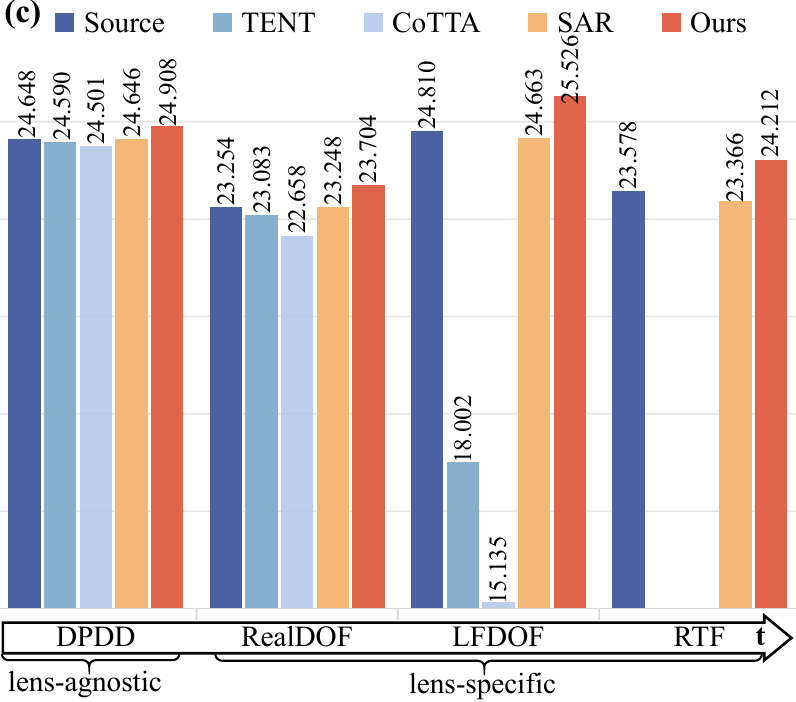}
\includegraphics[width=0.237\linewidth]{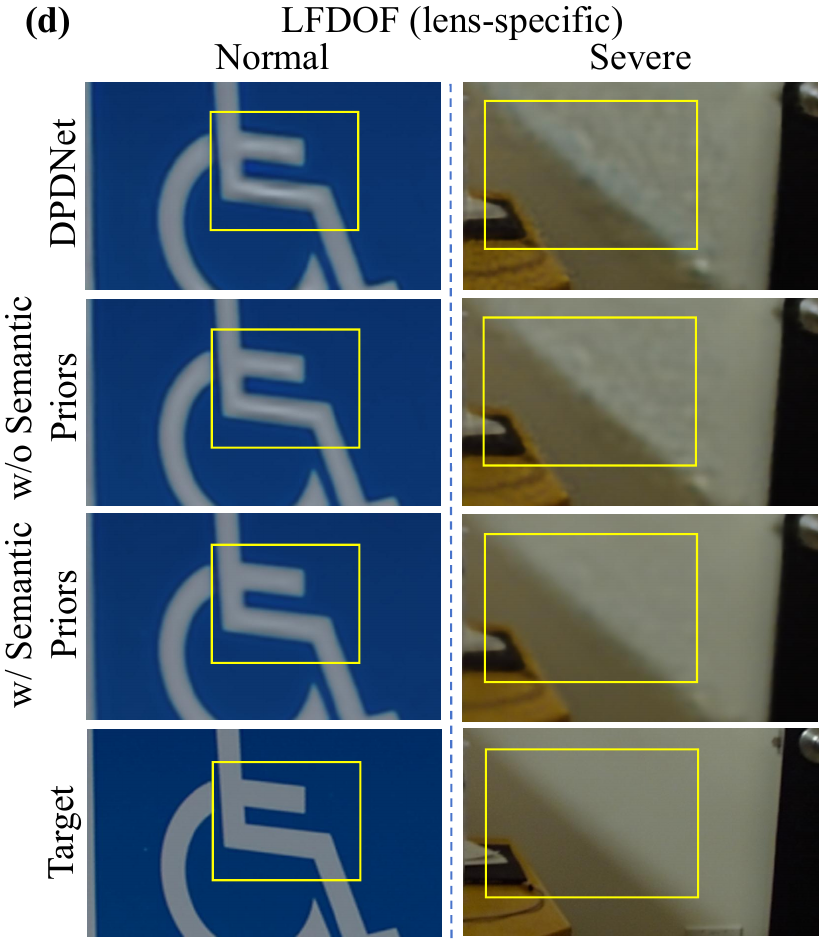}
\caption{Motivation experiments. (a) The visualization illustrates lens-specific and lens-agnostic PSF heterogeneity for two devices: Canon EOS 5D and Lytro Illum \citep{ruan2022learning}. (b) The NRKNet model, trained on the DPDD training set, successfully restores DPDD test images from the same device with the training set (lens-agnostic), but fails on LFDOF images from different devices (lens-specific), introducing false white artifacts. (c) Limited performance of existing CTTA algorithms (e.g., ``TENT", ``CoTTA", and ``SAR") during online continual adaptation over time. ``Source" represents the DPDNet-S \citep{abuolaim2020defocus} model trained on the DPDD dataset without adaptation. PSNR(dB) is used as the evaluation metric. (d) In cases of severe degradation, deblurring results without semantic priors exhibit semantically erroneous textures.}
\label{fig:movitation}
\end{figure*}

{The intuitive approach of SIDD is to estimate the defocus map and apply non-blind deconvolution for deblurring in a step-by-step manner \citep{levin2007image,d2016non,ma2021defocus}. However, this approach struggles with inaccurate blur kernel estimation and inefficient inference incurred by heavy computational demands. Recent deep neural network-based approaches \citep{abuolaim2020defocus,quan2023neumann} have demonstrated impressive performance using public paired datasets in an end-to-end manner. Nevertheless, their generalizability encounters challenges in out-of-distribution scenarios, due to the \textit{distribution shifts} between training and testing data \citep{ruan2022learning}.}
In practice, distribution shifts in the field of SIDD are characterized by two decoupled issues related to the point spread function (PSF\footnote{PSF describes the imaging system's response to a point source, representing how the point source of light is `spread' and characterizing the system's blurring effects \citep{goodman2005introduction}.}): 1) \textit{lens-agnostic PSF heterogeneity}, which refers to PSF heterogeneity caused by varying depth of field within the same optical system, i.e., the lens-related defocus characteristics are shared, while the depth of field varies; 2) \textit{lens-specific PSF heterogeneity}, which refers to PSF heterogeneity arising from different optical systems, i.e., lens-related defocus characteristics vary. Fig. \ref{fig:movitation}(a) provides visual examples that clarify these types of heterogeneity.

From the perspective of PSF heterogeneity, we conduct motivation experiments to explore the focal reason for the performance degradation of benchmark SIDD methods in out-of-distribution scenarios. As shown in Fig. \ref{fig:movitation}(b), the state-of-the-art (SOTA) method NRKNet \citep{quan2023neumann} effectively restores all-in-focus images from defocused images with respect to lens-agnostic PSF heterogeneity. However, it fails on lens-specific ones. {This highlights the potential limitation of existing SIDD algorithms, which primarily focus on lens-agnostic PSF heterogeneity and neglect lens-specific ones. This issue is attributed to the finiteness of training data and the lack of efficient mechanisms for adapting test data. In contrast, our proposed method leverages a nuanced design to underpin the generalization for lens-specific PSF heterogeneity. This motivation experiment further concludes that lens-specific PSF heterogeneity is a crucial factor contributing to the generalization degradation of benchmark SIDD approaches.}   

To tackle lens-specific PSF heterogeneity, we employ continual test-time adaptation (CTTA) \citep{wang2022continual}, which empowers trained models to adapt to varying inputs during inference. However, most CTTA algorithms are designed for classification tasks and rely on the entropy minimization theorem \citep{grandvalet2004semi}, which is unsuitable for pixel-wise regression tasks like SIDD. Directly applying existing CTTA algorithms to SIDD models fails to enhance their adaptation and may even further dilute their generalizability. {As illustrated in Fig. \ref{fig:movitation}(c), empirical evidence supports this analysis. The experiment result shows the performance of various CTTA algorithms on the continuous adaptation task involving four benchmark datasets with distinct distributions. For lens-agnostic test sets, the performance of the source SIDD model is comparable to that of the CTTA methods. However, for lens-specific test sets, these CTTA methods result in a significant drop in performance and even collapse. As time progresses (t), the likelihood of performance collapse increases. To overcome the challenge, we develop a novel CTTA learning paradigm that adapts SIDD models to new and unseen devices using limited data during the test phase. In this practical and constrained scenario, we achieve the goal by designing Siamese networks based on a consistency loss.}

However, when dealing with severely degraded lens-specific images, Siamese networks still struggle to eliminate semantically erroneous textures introduced by source models, as illustrated in the second row of Fig. \ref{fig:movitation}(d). To elucidate the intrinsic mechanism behind this empirical failure, we theoretically revisit the learning paradigm of SIDD from a causality perspective. Without loss of generality, we advance a structural causal model (SCM) \citep{pearl2009causality} to demonstrate the causal relationships between factors within SIDD. According to rigorous analysis of the proposed SCM, we confirm that the causal effect between the blurry and restored image is unidentifiable within only Siamese networks, which necessitates semantic knowledge. Guided by the SCM, we propose \textbf{Cau}sal \textbf{Siam}ese networks, dubbed \textbf{\textit{CauSiam}}, which integrate universal semantic priors derived from large-scale pre-trained vision-language models (VLMs) \citep{radford2021learning} into the Siamese networks. {VLMs are trained jointly on massive image-text pairs, enabling their image encoders to effectively map visual features to the semantic information from text. Concretely, they can capture rich and extensive semantic concepts from images, including objects, scenes, and situations. Thus, CauSiam empowers the SIDD model to simultaneously capture pixel-level and semantic-level features, guaranteeing the causal identifiability between blurry inputs and restored images.} Our main \textbf{contributions} are four-fold:
\begin{itemize}
\item We elaborate on the intrinsic reason exacerbating the performance degradation of SIDD models under out-of-distribution scenarios from the perspective of PSF heterogeneity, which is supported by sufficient motivation experiments.
\item We introduce the first CTTA framework for SIDD, mitigating performance degradation caused by distribution shifts stemming from lens-specific PSF heterogeneity between training and continually changing testing data.
\item To further enhance the performance of SIDD on severely degraded images and prevent semantically erroneous textures, we revisit the conducted learning paradigm through SCM-based theoretical analysis. Accordingly, we propose CauSiam, incorporating universal semantic priors derived from VLMs, to guarantee the causal identifiability between blurry inputs and restored images.
\item Extensive experiments demonstrate that the CauSiam framework improves the effectiveness and generalization of source SIDD models, as evidenced by their performance across five different SIDD test datasets.
\end{itemize}

\section{Related Work}
\label{section:relatedwork}
In this section, we comprehensively review relevant literature on single image defocus deblurring, continual test-time adaptation, and large-scale pre-trained vision-language models.

\begin{table*}[htbp]
\centering
\caption{The disparities among different adaptation algorithms encompass the available data, loss function, and distribution discrepancy during training and testing phases.}
\resizebox{\textwidth}{!}{
\begin{tabular}{lccccc}
\toprule[1pt]
\textbf{Setting} & \textbf{Source Data} & \textbf{Target Data} & \textbf{Train Loss} & \textbf{Test Loss}  & \textbf{Distribution}\\
\midrule
domain adaptation & $x_s, y_s$ & $x_t$ & $L(x_s, y_s) + L(x_s, x_t)$ & - &stationary\\
test-time training & $x_s, y_s$ & $x_t$ & $L(x_s, y_s) + L(x_t)$ & - &stationary\\
test-time adaptation & - & $x_t$ & - & $L(x_t)$ & stationary\\
continual test-time adaptation & - & $x_t$ & - & $L(x_t)$ & continually changing\\
\bottomrule[1pt]
\end{tabular}}
\label{tab:tta-settings}
\end{table*}

\subsection{Single Image Defocus Deblurring}
Single image defocus deblurring (SIDD) methods can be roughly divided into two categories: two-stage \citep{levin2007image,d2016non,li2019blind,liu2020estimating,ruan2021aifnet,ma2021defocus} and end-to-end deep learning-based approaches \citep{abuolaim2020defocus,lee2021iterative,son2021single,quan2021gaussian,ruan2022learning,zhai2023learnable,quan2023neumann,li2023learning,quan2024deep,tang2024prior}.

Two-stage methods estimate the defocus map and use non-blind deconvolution \citep{fish1995blind} to predict the sharp image. However, these methods are time-consuming and yield poor performance due to inaccurate defocus maps and over-idealized blur kernel models. The first end-to-end deep learning-based method \citep{abuolaim2020defocus} outperforms traditional two-stage approaches, but still struggles with spatially varying and large blurs. \citep{son2021single} address spatially varying blur by multiple kernel-sharing atrous convolutions and attention maps. \citep{quan2021gaussian,quan2024deep} develop a Gaussian kernel mixture (GKM) model that leverages the isotropy of defocus PSFs. \citep{lee2021iterative} adopt an iterative filter adaptive network trained with an extra reblur loss, while \citep{ruan2022learning} design a training strategy to adopt an extra light field dataset. \citep{zhao2022united} develop a weakly supervised framework for simultaneous defocus detection and deblurring. \citep{zamir2022restormer} propose an effective Transformer model for image restoration tasks, achieving outstanding results in SIDD and other tasks. \citep{quan2023neumann} propose a learnable recursive kernel representation for defocus kernels. \citep{tang2024prior} propose a prior-and-prediction inverse kernel transformer (P$^2$IKT), which comprises an inverse Gaussian kernel module and an inverse kernel prediction module.

However, these methods assume that the training and testing data come from the same distribution. In motion deblurring, meta-learning \citep{finn2017model} is commonly adapted to solve distribution shifts via test-time training \citep{chi2021test,liu2022meta}, optimizing meta-auxiliary tasks on source data before adapting to the target domain. Existing approaches do not cover all practical cases when source, target, or supervision data are not available simultaneously. In contrast, our work alleviates this burden without modifying the architecture, proxy tasks, source data, or training. 

\subsection{Continual Test-time Adaptation}
Based on test-time adaptation (TTA) \citep{wang2021tent,zhao2022delta,zhang2022memo,niu2022efficient,zhang2023domainadaptor,roy2023test,lee2024entropy}, trained models adapt to test data (target domain) in a source-free and online manner. As illustrated in Table \ref{tab:tta-settings}, TTA only requires the trained model and unlabeled target data during inference. TENT \citep{wang2021tent} adapts the affine parameters of batch normalization layers via entropy minimization to reduce generalization error on shifted data. EATA \citep{niu2022efficient} proposes an active sample identification strategy to select reliable and non-redundant samples for efficient TTA. Besides, several low-level TTA methods have been proposed. For instance, \citep{ren2020video} employs sharp frames from a video and proposes a fitting-to-test-data pipeline for video deblurring. SRTTA \citep{deng2023efficient} introduces a second-order degradation scheme as the self-supervised loss for real-world image super-resolution. While effective in mitigating domain gaps, TTA assumes the target domain maintains a static distribution. In the real world, target domain distributions are non-stationary and continuously changing over time, leading to error accumulation and catastrophic forgetting. 

To address these challenges, CTTA \citep{wang2022continual,niu2023towards,yuan2023robust,dobler2023robust,gan2023decorate,song2023ecotta,liu2024vida} comes into being as a result of the times. In the SIDD task, the target domain distribution continually changes due to different photography devices, making CTTA more suitable for SIDD than TTA. CoTTA \citep{wang2022continual} employs weight-averaged and augmentation-averaged predictions for refining pseudo labels. \citep{niu2023towards} propose a sharpness-aware and reliable entropy minimization method in the dynamic wild world. \citep{gan2023decorate} introduce image-level visual domain prompts for target domains to prevent catastrophic forgetting. VIDA \citep{liu2024vida} leverages high-rank and low-rank features for extracting domain-specific and domain-shared knowledge. Although CTTA methods are increasingly utilized in image classification \citep{song2023ecotta,gan2023decorate} and segmentation \citep{chen2024each,valanarasu2024fly}, they mostly rely on the entropy minimization theorem \citep{grandvalet2004semi} and cannot sufficiently explore task-dependent information for pixel-level regression tasks like SIDD. In this paper, we propose the first CTTA framework for SIDD and design Siamese networks by exploiting a consistency loss.

\subsection{Large-scale Pre-trained Vision-language Models}
Recently, a series of vision-language models pre-trained on large-scale datasets have emerged, such as CLIP \citep{radford2021learning}, DINOv2 \citep{oquab2024dinov}, and BLIP2 \citep{li2023blip}. These models provide rich and valuable knowledge for various high-level tasks like zero-shot classification \citep{zhai2022lit,esmaeilpour2022zero,gao2024clip}, image editing \citep{avrahami2022blended,patashnik2021styleclip}, open-world segmentation \citep{wang2022cris,zhou2023zegclip} and action recognition \citep{wang2023clip}. Recent studies \citep{wang2023exploring,liang2023iterative} also show that the visual-language priors encapsulated in CLIP can assess the image quality in a zero-shot manner. \citep{luo2024controlling} leverage CLIP to predict degradation and clean content embeddings for universal image restoration. \citep{xu2024boosting} leverage pre-trained models to enhance the performance of restoration tasks. \citep{yang2024ldp} estimate defocus blur maps from dual-pixel pairs in an unsupervised manner via CLIP. Recent studies leverage VLMs to enhance image restoration performance in supervised learning. However, our work focuses on improving the generalization of SIDD online under continuously changing distributions.

\begin{figure*}
	\centering
        \includegraphics[width=\textwidth]{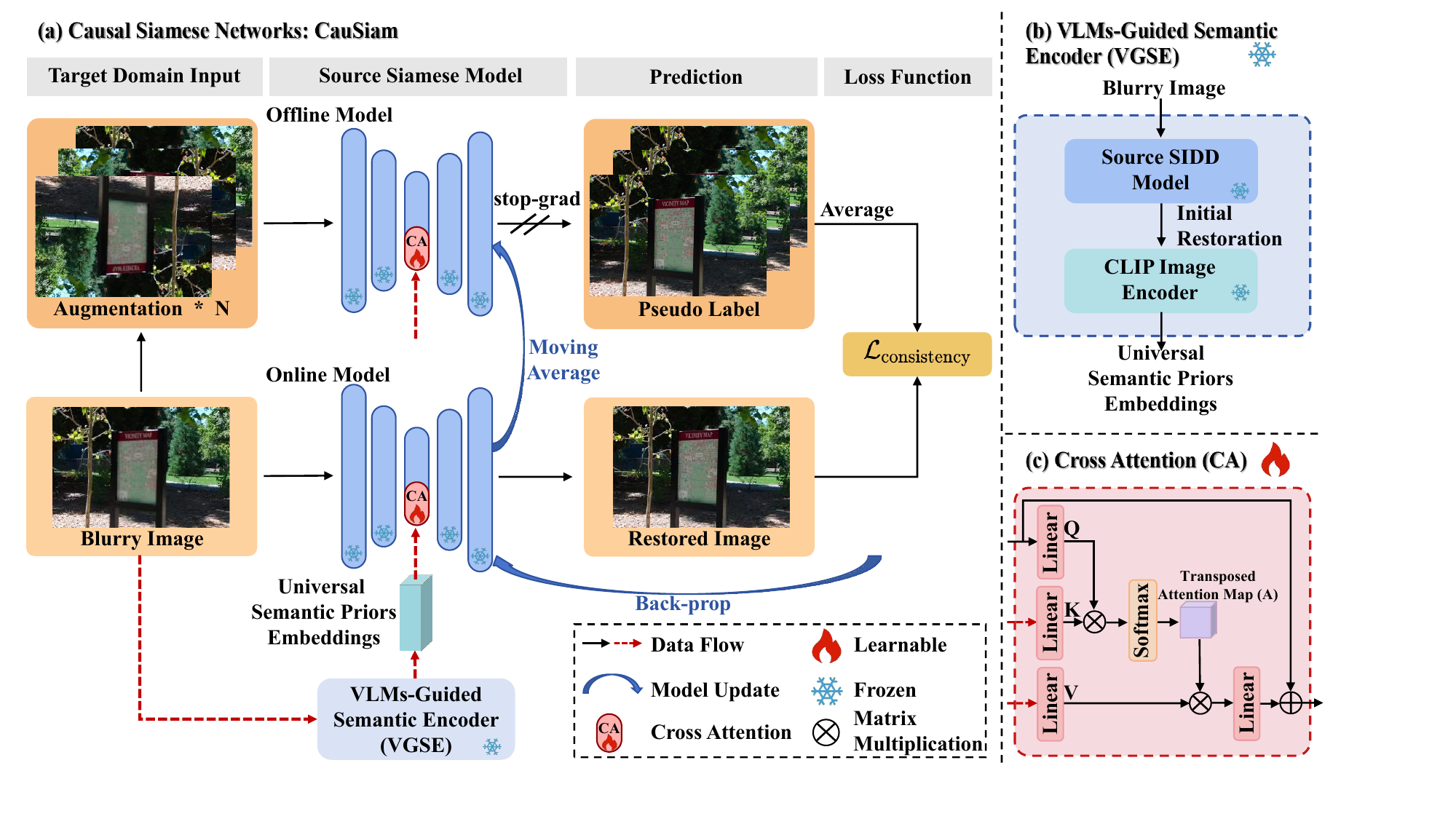}
	\caption{Framework of the proposed CauSiam. (a) The online model processes the original blurry image, while the offline model handles geometric augmentations (i.e., rotate and flip) to generate pseudo labels. We use a consistency loss (Equation \eqref{equ:loss_total}) as the optimization objective to update CauSiam. (b) VLMs-guided semantic encoder (VGSE) module extracts universal semantic priors embeddings for each test blurry image. (c) Cross attention (CA) module integrates these embeddings into the source SIDD model.}
	\label{fig:ssid-ctta}
\end{figure*}

\section{Methodology}
Here, we begin by introducing the problem definition of CTTA for SIDD in \textbf{Section} \ref{subsection:problem-definition}, and present the overall framework of our proposed CauSiam in \textbf{Section} \ref{subsection:framework}.

\subsection{Problem Definition}
\label{subsection:problem-definition}
In CTTA for SIDD, we adapt a source model $\mathcal{F}_\zeta$, trained on the source domain $\mathcal{D}_s=\left\{(x, y) \sim p_s(x, y)\right\}$, to multiple changing target domains $\mathcal{D}_t=\left\{(x) \sim p_t(x)\right\}$ online, where $x$ represents the defocused input and $y$ corresponds to the ground truth sharp image. The source data and target labels are inaccessible, and only the unlabeled target domain data can be accessed once. Notably, the source and target distributions differ ($p_s(x) \neq p_t(x)$), and the target domain distributions ($p_t(x)$) change over time. Our objective is to optimize the source model $\mathcal{F}_\zeta$ based on the current test sample $x_{\text{test}}^t$ at time t through an unsupervised loss $\mathcal{L}$ : 
\begin{equation} 
\zeta^*=\underset{\zeta}{\operatorname{argmin}} \mathcal{L}(\mathcal{F},\zeta,x_{\text{test}}^t) .
\label{equ:L_un}
\end{equation}

\subsection{Overview of the Proposed CauSiam}
\label{subsection:framework}

\begin{algorithm}[h]
\DontPrintSemicolon
  \SetAlgoLined
  \KwIn {A source SIDD model $\mathcal{F}_\zeta$; for time step $t$, the current single test sample $x_{\text{test}}^t$; hyper-parameters (e.g., iteration number $K$).}
  \KwOut{Adapted prediction $y_{\text{mean}}$; updated online model $\mathcal{F}_{\theta}$; updated offline model $\mathcal{F}_\xi$.}

  \For{$i$=1 to $K$}{
  Integrate causality-driven semantic priors into the source SIDD model to generate the online model $\mathcal{F}_\theta$ and the offline model $\mathcal{F}_\xi$. Initialize these models using the source SIDD model $\mathcal{F}_\zeta$;

  Augment the test sample and obtain the averaged pseudo label $y_{\text{mean}}$ from the offline model $\mathcal{F}_\xi$ by Equation \eqref{equ:avg-pseudo-labels};

  Update the online model $\mathcal{F}_{\theta}$ using the consistency loss in Equation \eqref{equ:loss_total};

  Update the offline model $\mathcal{F}_\xi$ by exponential moving average in Equation \eqref{equ:ema}.
  }

  \caption{CauSiam algorithm for SIDD during CTTA.}
  \label{alg:causiam}
\end{algorithm}

{Fig. \ref{fig:ssid-ctta} provides an overview of the CauSiam framework, and Algorithm \ref{alg:causiam} outlines the process. Overall, CauSiam consists of two key components, i.e., Siamese networks-based continual test-time adaptation (SiamCTTA) and causality-driven semantic priors integration (CSPI). The SiamCTTA module is designed to adapt source SIDD models to new and unseen devices, especially addressing the distribution shift caused by lens-specific PSF heterogeneity. Within this module, we utilize the online model $\mathcal{F}_{\theta}$ and the offline model $\mathcal{F}_{\xi}$, both initialized from the source SIDD model $\mathcal{F}_\zeta$. $\mathcal{F}_{\theta}$ and $\mathcal{F}_{\xi}$ share identical architectures but differ in their parameters. Additionally, the CSPI module is designed based on SCM theoretical analysis and integrates universal semantic priors from VLMs into source Siamese networks. This integration aims to guarantee the causal identifiability between blurry inputs and restored images. SiamCTTA and CSPI will be elaborated upon in detail in \textbf{Section} \ref{section:Siamese-CTTA} and \textbf{Section} \ref{section:causality-semantic-priors}, respectively.}

\section{Siamese Networks-Based Continual Test-Time Adaptation}
\label{section:Siamese-CTTA}
In this section, we will introduce the loss function and knowledge transfer mechanism of SiamCTTA.
\subsection{Loss Function}
\label{subsection:loss_function}
The primary challenge in CTTA lies in designing the test-time optimization objective that aligns with the characteristics of the SIDD task. When a test instance $x_{\text{test}}^t$ arrives at time $t$, we update parameters $\theta$ of the online SIDD model by optimizing an unsupervised loss function, as labels are not available for CTTA. For convenience, we abbreviate $x_{\text{test}}^t$ as $x_{\text{test}}$. Concretely, we develop augmentation consistency loss, including spatial consistency loss and high-frequency consistency loss.

\textbf{Spatial consistency loss.}
This objective promotes consistency in the model's predictions across different augmented views of a test image. Upon receiving a target domain input image, the models apply $N$ geometric augmentations. These augmentations involve inverse transformations to prevent information loss, specifically including horizontal and vertical flipping, as well as rotations of $90^{\circ}$, $180^{\circ}$, and $270^{\circ}$ clockwise. The offline model $\mathcal{F}_{\xi}$ processes these augmented images, producing $N$ deblurred outputs $\mathcal{F}_{\xi}(x_{\text{test}}^1)$, $\dots$, $\mathcal{F}_{\xi}(x_{\text{test}}^N)$. Then, these outputs are inverse-transformed back to their original geometry and averaged to create the refined pseudo label $y_{\text {mean}}$. Finally, we calculate the $\mathcal{L}_1$ regularization loss between the output of the online model $\mathcal{F}_{\theta}$ and the pseudo label $y_{\text {mean}}$, as follow:
\begin{equation}
\begin{aligned} \label{equ:avg-pseudo-labels}
\mathcal{L}_{\text{spatial}} &= \left\| \mathcal{F}_{\theta}\left(x_{\text{test}}\right) - y_{\text{mean}} \right\|_1, \\
y_{\text{mean}} &= \frac{1}{N} \sum_{n=1}^N \mathcal{F}_{\xi}\left(x_{\text{test}}^n\right),
\end{aligned}
\end{equation}
where $\mathcal{F}_{\theta}(x_{\text{test}})$ is the output of the online model, and $N$ is the number of augmentations. 

\textbf{High-frequency consistency loss.}
High-frequency components are essential for mitigating defocus blur, primarily impacting edges and fine details of images. We leverage the 2D Discrete Wavelet Transform (DWT) \citep{zhang2019wavelet} to decompose an image into four bands: approximation (LL), vertical (LH), horizontal (HL), and diagonal (HH). To manage the high-frequency components more effectively, we define $f^h(\cdot)$ as an operator that extracts and averages the high-frequency bands (LH, HL, HH). To mitigate the effect of defocus on high-frequency components, we minimize the high-frequency consistency loss, defined as:
\begin{equation} 
\mathcal{L}_{\text{high-frequency}} = \left\|f^h(\mathcal{F}_{\theta}\left(x_{\text{test}})\right)- f^h(y_{\text {mean }})\right\|_1.
\label{equ:loss_high-fre_consistency}
\end{equation}
By preserving high-frequency information, our method enhances image clarity by retaining edges and fine details.

\textbf{Full objective function.} The full objective function, augmentation consistency loss $\mathcal{L}_{\text{consistency}}$, for the online model is defined as:
\begin{equation} 
\mathcal{L}_{\text{consistency}}= \mathcal{L}_{\text{spatial}} + \lambda \mathcal{L}_{\text{high-frequency}},
\label{equ:loss_total}
\end{equation}
where $\lambda$ is hyper-parameter to balance the influence of $\mathcal{L}_{\text{spatial}}$ and $\mathcal{L}_{\text{high-frequency}}$.

\subsection{Knowledge Transfer Mechanism}
CTTA methods encounter challenges such as catastrophic forgetting and error accumulation due to the inherent noise and unreliability of pseudo labels \citep{wang2022continual}. To boost the stability of pseudo labels during adaptation, the offline model's parameters $\xi$ are updated as an exponential moving average (EMA) of the online model's parameters $\theta$. This mechanism allows the target domain knowledge obtained from the online model to be progressively distilled into the offline model, improving the accuracy of the pseudo label and fostering mutual learning between the models. Specifically, the update is performed after each test-time adaptation step:
\begin{equation} 
\xi \leftarrow (1-\eta) \xi + \eta \theta,
\label{equ:ema}
\end{equation}
where $\eta$ is the decay rate, and $\eta$ $\in$ [0,1].

\section{Causality-Driven Semantic Priors Integration}
\label{section:causality-semantic-priors}
SiamCTTA, proposed in \textbf{Section} \ref{section:Siamese-CTTA}, improves the generalizability of the source SIDD model while adapting to continually changing test data. However, as discussed in \textbf{Section} \ref{section:introduction}, when dealing with severely degraded images, it still cannot effectively eliminate semantically erroneous textures introduced by the source model. To profoundly comprehend the intrinsic mechanism of SIDD via SiamCTTA, we investigate the learning paradigm of the SIDD task from a causal perspective \citep{pearl2009causality}. The analysis is elucidated in the following \textbf{Section} \ref{subsection:theoretical_SCM}. Guided by the planned SCM, we integrate VLMs-guided semantic priors into SiamCTTA and propose CauSiam in \textbf{Section} \ref{subsection:vlms-semantic-priors}.

\subsection{Theoretical Analysis via Structural Causal Model}
\label{subsection:theoretical_SCM}
\begin{figure}
	\centering
        \includegraphics[width=0.45\textwidth]{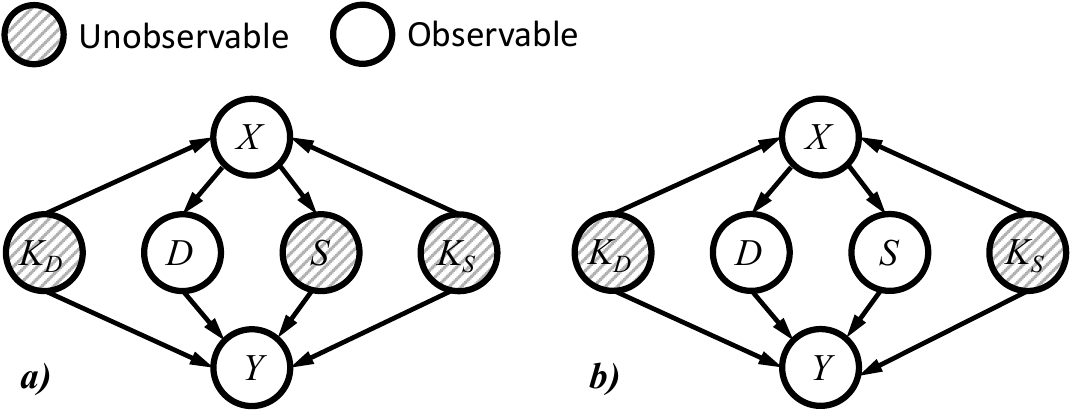}
	\caption{The proposed SCMs for SIDD. (a) The SCM of SIDD involves only SiamCTTA. (b) The SCM of SIDD regards the proposed CauSiam.}
	\label{fig:scm}
	\vspace{-0.2cm}
\end{figure}

{Firstly, we concisely describe the vanilla process of SiamCTTA. Given a blurry image, we obtain the corresponding $N$ augmentation images, then we leverage the consistency loss to update Siamese networks. Subsequently, we utilize the updated offline model to infer the final restored image. According to the depicted process, we propose the structural causal model (SCM) illustrated in Fig. \ref{fig:scm}(a), which holds for the following reasons:}

\begin{itemize}

\item \bm{$X \to D \to Y$} and \bm{$X \to S \to Y$}. $X$ is the blurry image to be restored. $D$ and $S$ denote the pixel-level feature and the semantic-level feature, respectively. The two components are preoccupied with distinct aspects, where $D$ inherently captures spatially variant details, while $S$ encompasses comprehensive semantic information. $Y$ is the restored image. We contend that $Y$ is jointly determined by $D$ and $S$ via two mediation ways, i.e., $X \to D \to Y$ and $X \to S \to Y$. To be specific, the reasons behind the above statement include: (\textbf{\romannumeral1}) $X \to D$ and $X \to S$, theoretically, based on the various well-designed encoders, we can extract the pixel-level feature $D$ and the semantic-level feature $S$ from the blurry image $X$; (\textbf{\romannumeral2}) $D \to Y$ and $S \to Y$, according to the attained pixel-level feature $D$ and semantic-level feature $S$, the restored image $Y$ can be derived. However, SiamCTTA primarily focuses on pixel-level content processing, thereby resulting in capturing the semantic-level feature by SiamCTTA is unachievable, which ultimately leads to the unobservable $S$.

\item \bm{$X \gets K_D \to Y$} and \bm{$X \gets K_S \to Y$}. $K_D$ and $K_S$ denote the complete pixel-level knowledge and the complete semantic-level knowledge, which are contained in $X$ (i.e., $K_D \to X \gets K_S$). Both $K_D$ and $K_S$ encompass task-independent information, therefore the direct ways $K_D \to Y$ and $K_S \to Y$ hold. Nonetheless, modeling the complete knowledge ($K_D, K_S$) is impractical: the knowledge modeling process is conventionally performed by leveraging the deep neural networks as encoders, while according to the data processing inequality \citep{beigi2013new}, the certain semantic knowledge degradation generally exists in the knowledge modeling process. Therefore, $K_D$ and $K_S$ are unobservable.
\end{itemize}

\begin{algorithm}[t] \label{alg:closedform}
\caption{ClosedForm($P(y|do(x))$.}
\KwIn{Control query of the form $P(y|do(x))$.}
\KwOut{Either a closed-form expression for $P(y|do(x))$, in terms of observed variables only, or \textcolor{myred}{FAIL} when the query is not identifiable.}
If ${(X \Vbar Y)}_{G_{\overline{X}}}$ then return $P(y)$\;

Otherwise, if ${(X \Vbar Y)}_{G_{\underline{X}}}$ then return $P(y|x)$\;

Otherwise, let $B=$ BlockingSet $(X,Y)$ and $Pb= \text{ClosedForm}( P( b\mid \hat{x} ) ) ;$ if
$Pb\neq$ \textcolor{myred}{FAlL} then return $\sum_{b}P(y\mid b,x)^*Pb$\;

Let $Z_{1}=\text{Children}(X)\cap(Y\cup\text{Ancestors}(Y)), Z_{3}$$=\text{BlockingSet}(X,Z_{1}),Z_{4}=$BlockingSet$(Z_{1},Y), \text{and}$
$
Z_{2}=Z_{3}\cup Z_{4}$; if $Y\notin Z_{1}\mathrm{~and~}X\notin Z_{2}$, then return $\sum\limits_{ z_{1},z_{2}}\sum\limits_{x^{\prime}}P(y\mid z_{1},z_{2},x^{\prime})P(x^{\prime}\mid z_2)P(z_{1}|x,z_{2})P(z_{2})$ \;

Otherwise, return \textcolor{myred}{FAIL}.

\end{algorithm}

Guided by the delineated SCM in Fig. \ref{fig:scm}(a), we attribute the degraded performance of SiamCTTA in handling severely degraded images to the unidentifiable causal effect between the blurry image and the restored image. It means that SiamCTTA can only fit the statistical correlation between $x$ and $y$ rather than the causal effect (i.e., $P(y|do(x))$). However, relying solely on the statistical correlation may distort or misguide the model's predictions. We confirm the unidentifiability of $P(y|do(x))$ by utilizing ClosedForm($\cdot$) function \citep{pearl2009causality}, which is elaborated in Algorithm \ref{alg:closedform}.  We systematically verify the validation of each step within this function in the following:
\textbf{(\romannumeral1)} Step 1, $G_{\overline{X}}$ denotes the SCM obtained by deleting all arrows pointing to the node $X$ in $G$, $X \Vbar Y$ means $X$ is independent of $Y$. Obviously, $(X \nVbar Y)_{G_{\overline{X}}}$ due to the existence of two directed paths (i.e., $X \to D \to Y$, $X \to S \to Y$).
\textbf{(\romannumeral2)} Step 2, $G_{\underline{X}}$ denotes the SCM obtained by deleting all arrows emerging from the node $X$ in $G$. $(X \nVbar Y)_{G_{\underline{X}}}$ due to the existence of two backdoor paths (i.e., $X \gets K_D \to Y$ and $X \gets K_S \to Y$); 
\textbf{(\romannumeral3)} Step 3, the function Blocking($X,Y$) selects a set of nodes $Z$ that $d$-separate $X$ from $Y$, and $Z=\{K_D,D,S,K_S\}$ in Fig. 3(a). However, $K_D,S,$ and $K_S$ are unobservable, leading to the condition in step 3 is not satisfied.;
\textbf{(\romannumeral4)} Step 4, $Z_1=\{D,S,Y\}$, $Y \in Z_1$ obviously and thus we can not calculate the causal effect by step 4;
\textbf{(\romannumeral5)} Step 5, due to the failure of steps 1-4, the function returns \textcolor{myred}{FAIL}, which means $P(y|do(x))$ is unidentifiable.

To tackle this dilemma and guarantee the causal identifiability between blurry inputs and restored images, the VLMs-guided semantic priors integration module is introduced to extract the semantic-level feature. With observable semantic-level feature $S$, SCM in Fig. \ref{fig:scm}(b) holds and we formalize the adjustment equation from the perspective of joint distribution decomposition \citep{pearl2009causality} in the following:

{\small \begin{subequations} \label{eq:front_door}
\begin{align} 
    & P(y|do(x)) \\
    & = \sum_{k_d} \sum_{k_s} \sum_d \sum_s P(k_d) P(k_s) P(d|x) P(s|x) P(y|k_d,k_s,d,s) \\ 
    & = \sum_{k_d} \sum_{k_s} \sum_d \sum_s P(k_d,k_s) P(d|x) P(s|x) P(y|k_d,k_s,d,s) \\
    & = \sum_{k_d} \sum_{k_s} \sum_d \sum_s \sum_{x'} P(k_d,k_s|x') P(x') P(d|x) P(s|x) \nonumber \\ 
    & \quad P(y|k_d,k_s,d,s) \\ 
    & = \sum_{k_d} \sum_{k_s} \sum_d \sum_s \sum_{x'} P(k_d,k_s|x',d,s) P(x') P(d|x) P(s|x) \nonumber \\ 
    & \quad P(y|k_d,k_s,d,s) \\
    & = \sum_{k_d} \sum_{k_s} \sum_d \sum_s \sum_{x'} \textcolor{myred}{ P(k_d,k_s|x',d,s) P(y|k_d,k_s,d,s,x')} \nonumber \\ 
    & \quad P(d|x) P(s|x) P(x') \\
    & = \sum_d \sum_s \sum_{x'} \textcolor{myred}{ P(y|d,s,x')} P(d|x) P(s|x) P(x'),
\end{align}
\end{subequations}}
where $x \in X$, $k_d \in K_D$, $k_s \in K_S$, $d \in D$, $s \in S$, and $y \in Y$.
Equation (\ref{eq:front_door}b) holds due to $K_D$ and $K_S$ are mutually independent; Equation (\ref{eq:front_door}c) and (\ref{eq:front_door}f) holds due to the law of total probability; Equation (\ref{eq:front_door}d) holds due to $K_D, K_S$ and $D, S$ are mutually independent given $X$; Equation (\ref{eq:front_door}e) holds due to $Y$ and $X$ are mutually independent given $K_D, D, S$, and $K_S$. Furthermore, we derive the adjustment equation from the perspective of multi-world symbolic derivation \cite{pearl2009causality}, which is consistent with Equation (\ref{eq:front_door}g). More details are provided in \textbf{Appendix} \ref{sec:appendix}.

\subsection{VLMs-Guided Semantic Priors Integration}
\label{subsection:vlms-semantic-priors}
 The above analysis concludes that the causal effect between the blurry and restored image is unidentifiable within only SiamCTTA, necessitating the introduction of semantic knowledge. To extract and integrate semantic knowledge of blurry images, we propose VLMs-guided semantic priors integration (VSPI), including VLMs-guided semantic encoder (VGSE) and cross attention (CA) module, as shown in Fig. \ref{fig:ssid-ctta}(b) and Fig. \ref{fig:ssid-ctta}(c).

\textbf{VLMs-guided semantic encoder (VGSE) module.} As illustrated in Fig. \ref{fig:ssid-ctta}(b), VGSE module comprises the frozen source SIDD model and pre-trained CLIP \citep{patashnik2021styleclip} image encoder. Extracting semantic information from defocused images is inherently challenging due to blur degradation. To mitigate the impact of blur, we first employ the source SIDD model to process the defocused image $x_{\text{test}}$, yielding the initial restoration $\mathcal{F}_{\zeta}(x_{\text{test}})$, which is subsequently resized to 224 $\times$ 224 pixels. Since CLIP is trained jointly on large-scale image-text pairs, its image encoder adeptly aligns image and language content. Concretely, CLIP image encoder can capture rich and extensive semantic content encapsulating comprehensive information from images, including object categories, scenes, and situations. Subsequently, we utilize the pre-trained CLIP image encoder to extract universal semantic priors embeddings $s_t$ from the resized initial restoration $\mathcal{F}_{\zeta}(x_{\text{test}})$:
\begin{equation} \label{eq:s_tclip}
s_t = \operatorname{ImageEncoder_{CLIP}}(\mathcal{F}_{\zeta}\left(x_{\text{test}}\right)).
\end{equation}

\begin{table*}
\begin{center}
\caption{Specifications of SIDD datasets.}
\begin{tabular}{lcccl}
\toprule[1pt] Dataset & Train Sample $\#$ & Test Sample $\#$ & Resolution  & Collect Device \\
\hline DPDD \citep{abuolaim2020defocus} & 350 & 76 & 1120 $\times$ 1680 &  Canon EOS 5D Mark IV \\
LFDOF \citep{ruan2021aifnet} & 11261 & 725 & 688 $\times$ 1008 & Lytro Illum camera\\
RTF \citep{d2016non} & None & 22 & 360 $\times$ 360 & Lytro camera \\
RealDOF \citep{lee2021iterative} & None & 50 & $\sim$ 1536 $\times$ 2320 & Sony $\alpha$ 7R IV \\
CUHK \citep{shi2014discriminative} & None & 704 & $\sim$ 470 $\times$ 610 & Internet \\
\toprule[1pt]
\end{tabular}
\label{table:dataset}
\end{center}
\end{table*}

\textbf{Cross attention (CA) module.} Since off-the-shelf features from VLMs are not directly applicable to image restoration \citep{xu2024boosting} like SIDD, we introduce CA module to dynamically integrate the semantic guidance feature and the pixel-level feature from the source SIDD model in a self-adaptative attention manner. Considering the increasing cost of applying attention to high-resolution features, we apply CA module only in the bottom block of the source SIDD models. Here, we define the feature from the bottom block of the source SIDD model as input intermediate features $z_s$. In CA module, input intermediate features $z_s$ are initially converted at the token level and then projected into the query ($\mathbf{Q}$) matrix. Universal semantic priors embeddings $s_t$ from VGSE module are projected into key ($\mathbf{K}$) and value ($\mathbf{V}$) matrices at token levels. As follows:
\begin{equation}
\begin{aligned}
\mathbf{Q} &= W_{\Theta^Q}\left(z_s\right) + B_{\Theta^Q}, \\
\mathbf{K} &= W_{\Theta^K}\left(s_t\right) + B_{\Theta^K}, \\
\mathbf{V} &= W_{\Theta^V}\left(s_t\right) + B_{\Theta^v},
\end{aligned}
\end{equation}
where $W_{\Theta^l}$ and $B_{\Theta^l}$ ($l \in \{\mathbf{Q}, \mathbf{K}, \mathbf{V}\}$) are projector and bias terms updated during testing, $z_s$ represents input intermediate features. After that, the semantic-aware attention map $\mathbf{A}$ can be formulated as: 
\begin{equation}
\mathbf{A}=\operatorname{Softmax}\left(\frac{\mathbf{Q K}^T}{\sqrt{C}}\right),
\end{equation}
where $(\cdot)^T$ represents the transpose operation, and $C$ is the channel of features. Finally, the predictions obtained through the attention fusion layer could be formulated as below:
\begin{equation}
z_s'=\alpha \mathbf{A} \mathbf{V} + z_s,
\end{equation}
where $z_s'$ represents output intermediate features after incorporating semantic priors, $\alpha$ is the hyper-parameter to balance the influence of semantic-level features and pixel-level intermediate features from the online model. For the offline model, CA module is utilized similarly.

During the back-propagation of the online model $\mathcal{F}_{\theta}$, only CA module is updated, while others remain frozen. Consequently, only the parameters of $W_{\Theta^l}$ and $B_{\Theta^l}$ are adjusted to accommodate the test-time distribution. This updating mechanism enhances the efficiency of CauSiam and preserves the knowledge of the source SIDD model, effectively mitigating catastrophic forgetting and error accumulation. 

\section{Experiments}
\label{section:experiments}
{In this section, we first outline the experimental settings and then present quantitative and qualitative experiment results. Next, we perform ablation studies to illustrate the effectiveness of each component. Finally, we conduct a comprehensive analysis of computational complexity.}

\subsection{Experimental Settings}
\label{subsection:experiments-setup}

\textbf{Datasets.} With CTTA training paradigm, we evaluate the performance of CauSiam and its counterparts on five public SIDD test sets, including DPDD \citep{abuolaim2020defocus}, RealDOF \citep{lee2021iterative}, LFDOF \citep{ruan2021aifnet}, RTF \citep{d2016non}, and CUHK \citep{shi2014discriminative}. We depict details of the datasets in Table \ref{table:dataset}, which shows that these datasets are collected from different devices with varying resolutions, thereby resulting in lens-agnostic and lens-specific PSF heterogeneity. Since the training set of source models includes the DPDD training set but excludes those of other datasets, distribution shifts encountered when adapting source models to the DPDD test set are attributed to lens-agnostic PSF heterogeneity. In contrast, distribution shifts for other datasets stem from lens-specific PSF heterogeneity.

\textbf{Compared methods.} Low-level TTA methods (Ren et al. \citep{ren2020video} and SRTTA \citep{deng2023efficient}) and low-level TTT methods (Chi et al. \citep{chi2021test}) serve as straightforward baselines for comparison. In addition, based on source models, we compare CauSiam with four competitive CTTA methods, including a pseudo-label-based method (CoTTA \citep{wang2022continual}) and three entropy-minimization-based methods (TENT-continual \citep{wang2021tent}, TENT-online* \citep{wang2021tent}, and SAR \citep{niu2023towards}). TENT-online* is the online version of TENT-continual, which resets the weights of the source model when encountering new target domains. To ensure a fair comparison, we replace the entropy-based consistency loss in CoTTA with $\mathcal{L}_1$ regularization consistency loss to better fit the characteristics of the pixel-level SIDD task. 

To comprehensively demonstrate the effectiveness and generalizability of CauSiam, we select representative source SIDD models: GGKMNet \citep{quan2024deep}, DPDNet-S \citep{abuolaim2020defocus}, IFANet \citep{lee2021iterative}, Restormer \citep{zamir2022restormer}, DRBNet \citep{ruan2022learning}, NRKNet \citep{quan2023neumann}, and P$^2$IKT \citep{tang2024prior}. With the exception of DRBNet, all these models were trained using the DPDD training set, as the original DRBNet paper only provides a model trained on the DPDD and LFDOF datasets. Due to the memory limitations of Restormer \citep{zamir2022restormer}, we randomly crop the entire image to a small part before processing. Since the code of GGKMNet is not publicly available, we employ it as a straightforward baseline for comparison, without incrementally incorporating CTTA methods. It is worth noting that the existing domain adaptation algorithms for SIDD, such as DMENet \citep{lee2019deep}, require ground truth defocus maps for training. However, because neither DPDD nor LFDOF datasets include ground truth defocus maps, we exclude these domain adaptation algorithms from our baselines.

\textbf{Evaluation metrics.} We primarily employ full-reference evaluation metrics and visualization results to evaluate the performance. Specifically, for the DPDD, RealDOF, LFDOF, and RTF datasets, we select common metrics such as Peak Signal-to-Noise Ratio (PSNR) and Structural Similarity Index (SSIM) \citep{wang2004image}. For the CHUK dataset, which lacks paired ground truth data, we rely exclusively on visualization results.

\textbf{Implementation details.} During the test phase, we perform an iteration $K$ to adapt each batch, we set the batch size to 1 across all experiments. In CA module, there is one layer consisting of 8 attention heads, each head with a dimensionality of 64. The channel number of the universal semantic priors embeddings $s_t$ is 512. To update CauSiam, we utilize Adam optimizer with $\beta_1$ = 0.9, $\beta_2$ = 0.99, and a learning rate of 1e-4. We choose ViT-B/32 \citep{dosovitskiy2020image} as the backbone of CLIP image encoder. Hyper-parameters $\alpha$ (weight of semantic-level features), $\eta$ (target decay rate of EMA), and $\lambda$ (weight of high-frequency consistency loss) are set to 0.05, 0.9, and 0.01, respectively. The number of geometric augmentations $N$ is set to 5. We implement CauSiam by PyTorch, and all experiments are conducted on an NVIDIA RTX 3090Ti GPU.

\begin{table*}
\caption{{Quantitative comparison of three low-level TTA or TTT, SIDD source baseline, four CTTA methods, and our CauSiam in the continual test-time adaptation SIDD task. * indicates the requirement for additional domain information. $\uparrow$ denotes that larger values lead to better quality. \textcolor{myred}{+} denotes the \textcolor{myred}{improvement} of performance. \textbf{Bold} denotes the best performance and \underline{underline} stands for the second-best performance.}}
\centering
\tabcolsep=0.07cm
\begin{tabular}{l|cc|cc|cc|cc|cc}
\toprule[1pt]
Time & \multicolumn{8}{c}{t}{$\xrightarrow{\hspace{8cm}}$} \\
\midrule
\multirow{2}{*}{Method} & \multicolumn{2}{c|}{DPDD} & \multicolumn{2}{c|}{RealDOF} & \multicolumn{2}{c|}{LFDOF} & \multicolumn{2}{c|}{RTF} & \multicolumn{2}{c}{Average} \\
& PSNR & SSIM & PSNR & SSIM& PSNR & SSIM& PSNR & SSIM& PSNR$\uparrow$ & SSIM$\uparrow$\\
\midrule[1pt]
{Ren et al.} \citep{ren2020video} & {23.686} & {0.725} & {22.259} & {0.629} & {25.459} & {0.765} &	{23.728} & {0.751} & {25.078} & {0.753} 
\\
{SRTTA} \citep{deng2023efficient} & {24.442} & {0.735} &	{22.376} & {0.640} & {24.739} & {0.757} & {23.990} & {0.740} & {24.559} & {0.748} 
\\
{Chi et al.} \citep{chi2021test} & {24.166} & {0.731} & {22.244} & {0.638} & {24.993} & {0.757} &	{23.556} & {0.718} & {24.727} & {0.747} 
\\
\midrule
{GGKMNet} \citep{quan2024deep} & {26.039} & {0.806} & {24.942} & {0.763} & {\textemdash} & {\textemdash} & {25.895} & {0.827} & {\textemdash} & {\textemdash}\\
{GGKMNet$^\dag$} \citep{quan2024deep} & {26.272} & {0.810} & {25.355} & {0.770} & {\textemdash} & {\textemdash} & {26.012} & {0.846} & {\textemdash} & {\textemdash}\\
\midrule
DPDNet-S \citep{abuolaim2020defocus} & 24.648 & 0.758 & 23.254 & 0.686 & 24.810 & 0.768 & 23.578 & 0.757 & 24.676 & 0.762 \\
+ CoTTA \citep{wang2022continual} & 24.633 & 0.754& 22.439 & 0.650 & 22.746 & 0.727 & 21.583 & 0.685  & 22.863 & 0.724 \\
+ TENT-continual  \citep{wang2021tent} & 24.590 & \underline{0.758} & 23.083 & 0.685 & 18.002 & 0.693 & 11.013 & 0.549  & 18.690 & 0.695\\
+ TENT-online*  \citep{wang2021tent} & 24.590 & \underline{0.758} & 23.240 & \underline{0.686} & 19.244 & 0.715 & \underline{23.569} & 0.757 & 20.047 & 0.718 \\
+ SAR \citep{niu2023towards} & \underline{24.646} & \underline{0.758} & \underline{23.248} & \underline{0.686} & \underline{24.663} & \underline{0.768} & 23.366 & \underline{0.758} & \underline{24.548} & \underline{0.762} \\
\rowcolor{grayblue}
+ CauSiam(Ours) & \textbf{24.862} & \textbf{0.759} & \textbf{23.713} & \textbf{0.687} & \textbf{25.617} & \textbf{0.778} & \textbf{24.399} & \textbf{0.785} & \textbf{25.412} & \textbf{0.774} \\
Ours Gains & \textcolor{myred}{+0.214} & \textcolor{myred}{+0.001} & \textcolor{myred}{+0.459} & \textcolor{myred}{+0.001} & \textcolor{myred}{+0.807} & \textcolor{myred}{+0.010} & \textcolor{myred}{+0.821} & \textcolor{myred}{+0.028} & \textcolor{myred}{+0.736} & \textcolor{myred}{+0.012} \\
\midrule
IFANet \citep{lee2021iterative} & 25.364 & 0.788 & 24.707 & 0.748 & 26.107 & 0.816 & 24.926 & 0.821 & 25.932 & 0.810 \\
+ CoTTA \citep{wang2022continual} & \underline{25.704} & \textbf{0.795} & \underline{25.102} & \underline{0.758} & \underline{26.434} & \underline{0.821} & \underline{25.489} & \underline{0.825} & \underline{26.270} & \underline{0.815} \\
+ TENT-continual  \citep{wang2021tent} & 25.313 & \underline{0.788} & 24.672 & 0.749 & 21.420 & 0.746 & 17.211 & 0.597  & 21.839 & 0.746 \\
+ TENT-online*  \citep{wang2021tent} & 25.313 & \underline{0.788} & 24.685 & 0.748 & 22.090 & 0.739 & 24.899 & 0.821& 22.590 & 0.746 \\
+ SAR \citep{niu2023towards} & 25.363 & \underline{0.788} & 24.707 & 0.748 & 26.077 & 0.816 & 24.888 & 0.821 & 25.907 & 0.810 \\
\rowcolor{grayblue}
+ CauSiam(Ours) & \textbf{25.756} & \textbf{0.795} & \textbf{25.204} & \textbf{0.762} & \textbf{26.661} & \textbf{0.825} & \textbf{25.804} & \textbf{0.837} & \textbf{26.478} & \textbf{0.819} \\
Ours Gains & \textcolor{myred}{+0.392} & \textcolor{myred}{+0.007} & \textcolor{myred}{+0.497} & \textcolor{myred}{+0.014} & \textcolor{myred}{+0.554} & \textcolor{myred}{+0.008} & \textcolor{myred}{+0.878} & \textcolor{myred}{+0.016} & \textcolor{myred}{+0.546} & \textcolor{myred}{+0.009} \\
\midrule
DRBNet \citep{ruan2022learning} & 25.722 & 0.791 & 25.743 & 0.770 & 27.737 & 0.836 & 26.221 & 0.853 & 27.409 & 0.829 \\
+ CoTTA \citep{wang2022continual} & \textbf{25.754} & \textbf{0.793} & \underline{25.864} & \underline{0.773} & \underline{27.795} & \underline{0.835} & \underline{26.121} & 0.837 & \underline{27.464} & \underline{0.828} \\
+ TENT-continual  \citep{wang2021tent} & 24.241 & \underline{0.757} & 22.247 & 0.696 & 26.258 & 0.809 & 25.606 & \underline{0.846} & 25.836 & 0.799 \\
+ TENT-online*  \citep{wang2021tent} & 24.241 & \underline{0.757} & 23.117 & 0.711 & 26.206 & 0.813 & 25.711 & 0.840 & 25.846 & 0.803 \\
+ SAR \citep{niu2023towards} & 24.237 & \underline{0.757} & 22.200 & 0.695 & 26.117 & 0.807 & 25.484 & 0.844 & 25.713 & 0.797 \\
\rowcolor{grayblue}
+ CauSiam(Ours) & \underline{25.750} & \textbf{0.793} & \textbf{25.865} & \textbf{0.777} & \textbf{28.085} & \textbf{0.843} & \textbf{26.476} & \textbf{0.861} & \textbf{27.714} & \textbf{0.835} \\
Ours Gains & \textcolor{myred}{+0.028} & \textcolor{myred}{+0.002} & \textcolor{myred}{+0.122} & \textcolor{myred}{+0.007} & \textcolor{myred}{+0.348} & \textcolor{myred}{+0.007} & \textcolor{myred}{+0.255} & \textcolor{myred}{+0.008} & \textcolor{myred}{+0.305} & \textcolor{myred}{+0.006} \\
\midrule
Restormer \citep{zamir2022restormer} & 25.977 & 0.810 & 25.086 & 0.771 & 26.438 & 0.823 & 24.236 & 0.823 & 26.265 & 0.819\\
+ CoTTA \citep{wang2022continual} & \underline{26.039} & 0.808 & 24.000 & 0.716 & 26.353 & 0.808 & \textbf{24.768} & 0.811 & 26.151 & 0.803\\
+ TENT-continual  \citep{wang2021tent} & 25.977 & 0.810 & 25.086 & \underline{0.771} & 26.429 & 0.823 & 24.230 & 0.822  & 26.258 & 0.819\\
+ TENT-online*  \citep{wang2021tent} & 25.977 & 0.810 & \underline{25.087} & \underline{0.771} & 26.432 & 0.823 & 24.235 & \underline{0.823} & 26.260 & 0.819\\
+ SAR \citep{niu2023towards} & 25.978 & \underline{0.811} & 25.086 & \underline{0.771} & \underline{26.468} & \underline{0.824} & 24.235 & \underline{0.823} & \underline{26.290} & \underline{0.820} \\
\rowcolor{grayblue}
+ CauSiam(Ours) & \textbf{26.146} & \textbf{0.815} & \textbf{25.328} & \textbf{0.780} & \textbf{26.598} & \textbf{0.828} & \underline{24.387} & \textbf{0.830} & \textbf{26.431} & \textbf{0.824}  \\
Ours Gains & \textcolor{myred}{+0.169} & \textcolor{myred}{+0.005} & \textcolor{myred}{+0.242}  & \textcolor{myred}{+0.009} & \textcolor{myred}{+0.160} & \textcolor{myred}{+0.005} & \textcolor{myred}{+0.151} & \textcolor{myred}{+0.007} & \textcolor{myred}{+0.166} & \textcolor{myred}{+0.005} \\
\midrule
NRKNet \citep{quan2023neumann} & 26.109 & 0.803 & 25.027 & 0.752 & 26.398 & 0.811 & 25.949 & 0.847 & 26.283 & 0.808 \\
+ CoTTA \citep{wang2022continual} & \underline{26.156} & \underline{0.806} & 24.939 & 0.747 & 25.873 & 0.809 & 23.649 & 0.832 & 25.788 & 0.806 \\
+ TENT-continual  \citep{wang2021tent} & 26.102 & 0.803 & 24.986 & 0.751 & 18.539 & 0.653 & 8.749 & 0.370 & 19.320 & 0.664 \\
+ TENT-online*  \citep{wang2021tent} & 26.102 & 0.803 & 25.067 & \underline{0.752} & 19.690 & 0.685 & \underline{25.906} & \underline{0.845} & 20.712 & 0.703 \\
+ SAR \citep{niu2023towards} & 26.116  & 0.803 & \underline{25.080} & \underline{0.752} & \underline{26.543} & \underline{0.814} & 25.825 & \underline{0.845} & \underline{26.404} & \underline{0.810}\\
\rowcolor{grayblue}
+ CauSiam(Ours) & \textbf{26.204} & \textbf{0.807} & \textbf{25.220} & \textbf{0.758} & \textbf{26.705} & \textbf{0.817} & \textbf{26.087} & \textbf{0.850} & \textbf{26.560} & \textbf{0.814}\\
Ours Gains & \textcolor{myred}{+0.095} & \textcolor{myred}{+0.004} & \textcolor{myred}{+0.193} & \textcolor{myred}{+0.006} & \textcolor{myred}{+0.307} & \textcolor{myred}{+0.006} & \textcolor{myred}{+0.138} & \textcolor{myred}{+0.003} & \textcolor{myred}{+0.277} & \textcolor{myred}{+0.006}\\
\midrule
{P$^2$IKT} \citep{tang2024prior} & {26.284}  & {0.807}  & {25.479}  & {0.762}  & {26.897}  & {0.820}  & {25.883}  & {0.840}  & {26.737}  & {0.816}   \\
{+ CoTTA} \citep{wang2022continual} & \underline{{26.344}}  & \underline{{0.809}}  & {25.360}  & \underline{{0.767}}  & {26.754}  & \underline{{0.822}}  & \underline{{25.929}}  & \textbf{{0.842}}  & {26.618}  & \underline{{0.818}} \\
{+ TENT-continual} \citep{wang2021tent} & {26.291}  & {0.807}  & {25.331}  & {0.761}  & {25.079}  & {0.808}  & {22.460}  & {0.814}  & {25.133}  & {0.805}  \\
{+ TENT-online*} \citep{wang2021tent} & {26.291}  & {0.807}  & {25.440}  & {0.761}  & {25.253}  & {0.811}  & {25.859}  & {0.840}  & {25.370}  & {0.809 } \\
{+ SAR} \citep{niu2023towards} & {26.285}  & {0.807}  & \underline{{25.472}}  & {0.762}  & \underline{{26.835}}  & \underline{{0.822}}  & {25.806}  & {0.840}  & \underline{{26.683}}  & \underline{{0.818}} \\
\rowcolor{grayblue}
{+ CauSiam(Ours)} & {\textbf{26.378}}  & {\textbf{0.810}}  & {\textbf{25.664}}  & {\textbf{0.771}}  & {\textbf{27.025}}  & {\textbf{0.825}}  & {\textbf{26.151}}  & {\textbf{0.842}}  & {\textbf{26.869}}  & {\textbf{0.821}} \\
{Ours Gains} & \textcolor{myred}{+0.094}  & \textcolor{myred}{+0.003} & \textcolor{myred}{+0.185}  & \textcolor{myred}{+0.009} & \textcolor{myred}{+0.128}  & \textcolor{myred}{+0.005}  & \textcolor{myred}{+0.268}  & \textcolor{myred}{+0.002}  & \textcolor{myred}{+0.132} & \textcolor{myred}{+0.005}  \\
\bottomrule[1pt]
\end{tabular}
\label{table:quantitative-comparison}
\end{table*}

\begin{figure*}
	\centering
        \includegraphics[width=\textwidth]{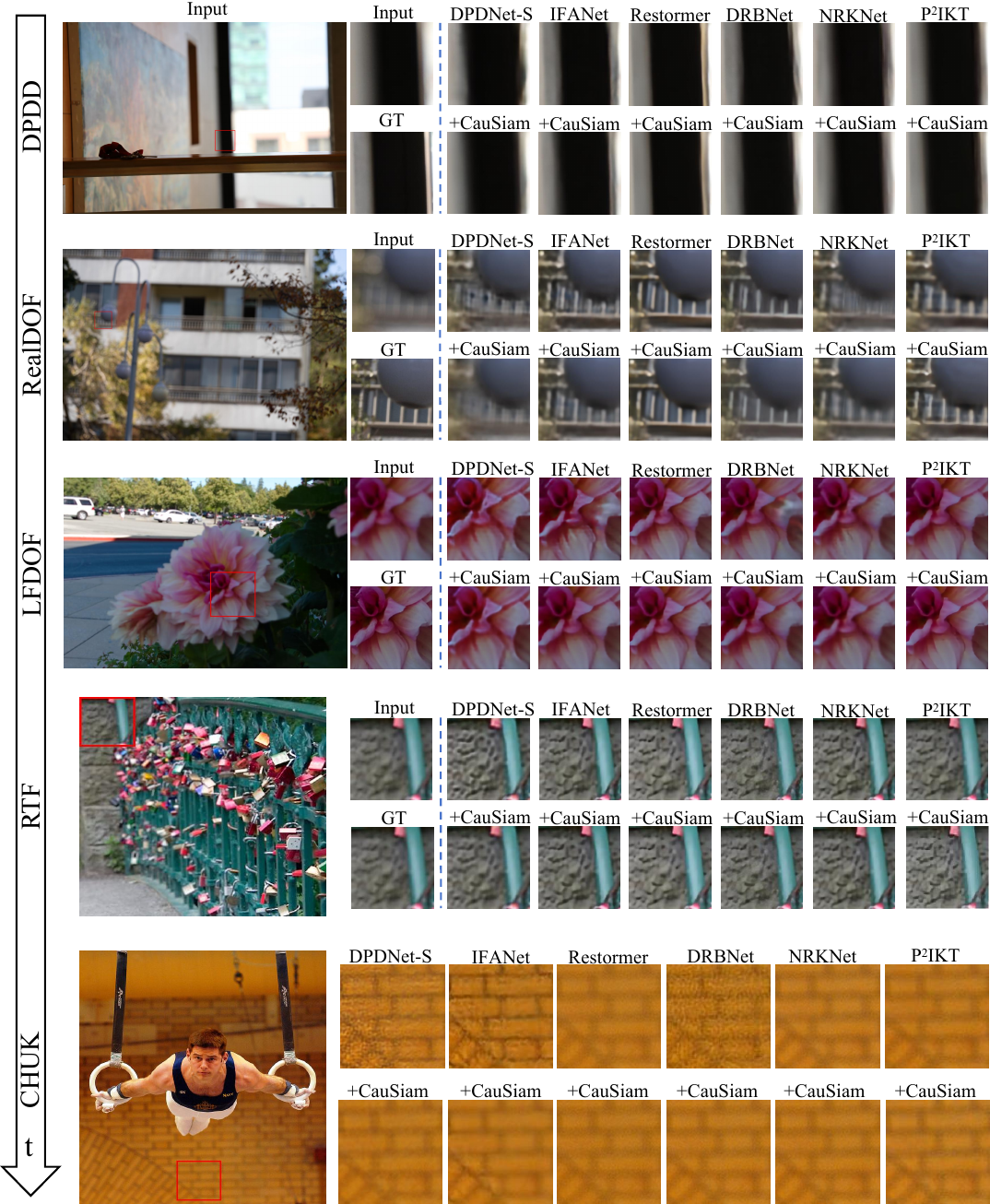}
	\caption{{Qualitative comparison of source SIDD models with and without our CauSiam on DPDD, RealDoF, LFDOF, RTF, and CHUK test sets during continual test-time adaptation.  The odd-numbered rows (1st, 3rd, 5th, 7th, and 9th) show visualizations of different source SIDD models trained on the DPDD training set without adaptation. The even-numbered rows (2nd, 4th, 6th, 8th, and 10th) display the visualizations after integrating CauSiam into source models.}}
	\label{fig:qualitative-exp-sidd}
\end{figure*}

\begin{figure*}
	\centering
    \includegraphics[width=\textwidth]{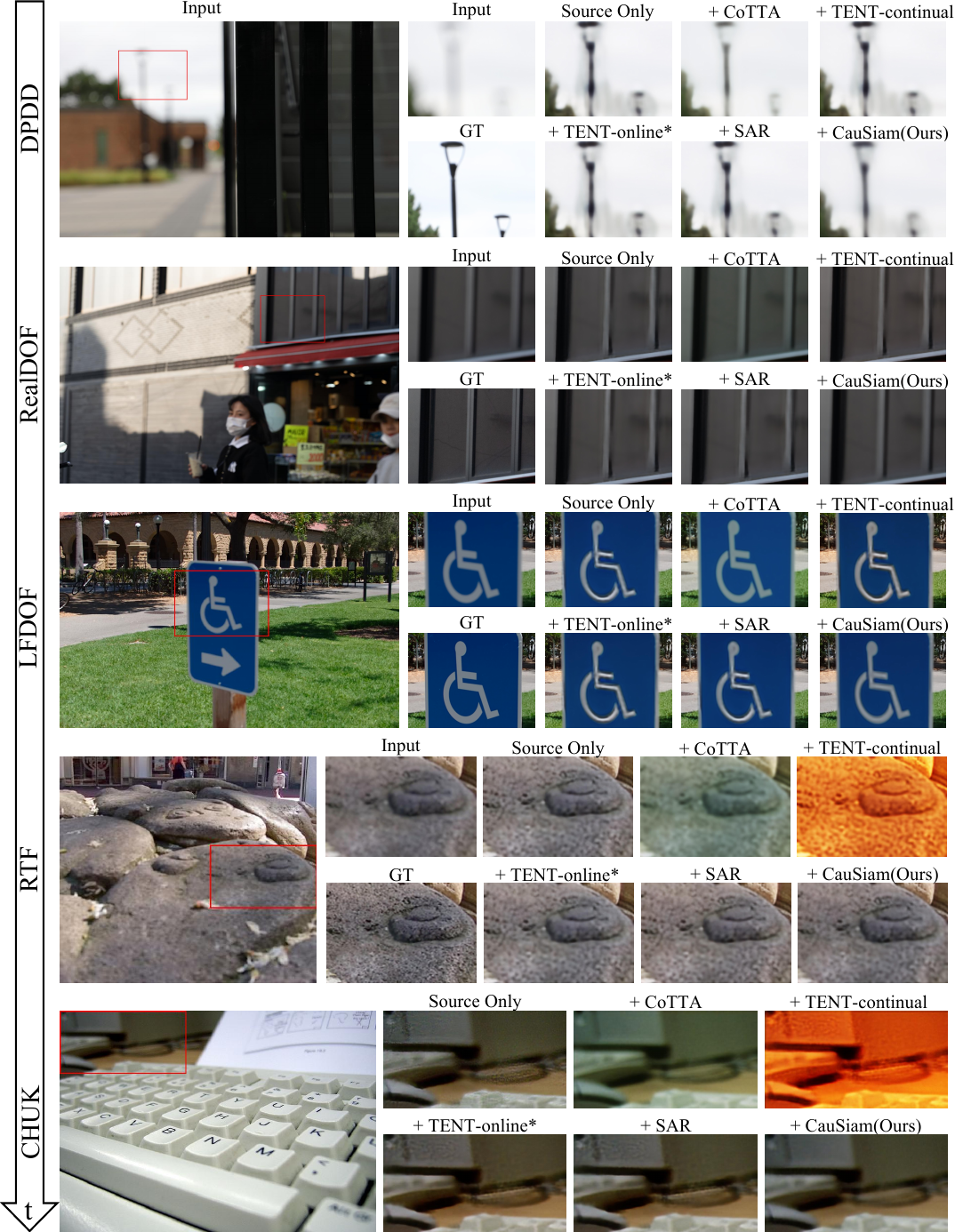}
	\caption{Qualitative comparison of different CTTA methods on DPDD, RealDoF, LFDOF, RTF, and CHUK test sets during continual test-time adaptation. ``Source Only" denotes the DPDNet-S model trained on the DPDD training set without adaptation.}
	\label{fig:qualitative-exp-ctta}
\end{figure*}

\subsection{Quantitative Comparison}

Table \ref{table:quantitative-comparison} presents the quantitative results of low-level TTA or TTT methods, SIDD source models, four CTTA methods, and our CauSiam in the continual test-time adaptation SIDD task.

Our proposed CauSiam demonstrates superior performance compared to low-level TTA and TTT methods, such as Ren et al., SRTTA, and Chi et al. While we retrain these methods using the same DPDD dataset, their self-supervised optimization objectives are tailored to their specific applications, rendering them less suitable for the SIDD task. Additionally, these methods do not account for continual adaptation, as distributions of the target domain are non-stationary and change over time.

When integrated with CauSiam, baseline SIDD source models achieve significant performance improvements, especially in restoring images degraded by lens-specific PSF heterogeneity. For instance, CauSiam enhances the source DPDNet-S model with PSNR/SSIM gains of 0.807 dB/0.010 on LFDOF and boosts IFANet by 0.497 dB/0.014 on RealDOF. The performance gains diminish as model capacity increases, yet they remain significant. For example, the average PSNR/SSIM gains of NRKNet are 0.307 dB/0.006 on LFDOF and 0.277 dB/0.006 across four SIDD datasets. DRBNet is trained on the combined training set of DPDD and LFDOF, which means there exists slighter distribution shifts. Nevertheless, ``DRBNet + CauSiam (Ours)" still shows an improvement of 0.348 dB/0.007 on LFDOF, suggesting the robust generalization capability of our algorithm. {While P$^2$IKT demonstrates strong generalization ability, integrating CauSiam significantly enhances its performance, resulting in an average PSNR improvement of 0.132 dB. Furthermore, ``P$^2$IKT + CauSiam (Ours)" surpasses the recent SOTA model GGKMNet$^\dag$.} We attribute the performance improvement of CauSiam to two reasons: 1) Leveraging Siamese networks for augmented consistency and VLMs for semantic priors, we reduce artifacts and preserve details. 2) The mechanism underlying CauSiam can conduct effective distribution alignment between the source domain and the target domain.

Compared to other competitive CTTA methods, CauSiam provides the most significant enhancement to source models. The experimental results indicate that entropy-minimization-based CTTA methods are not suitable for pixel-level regression tasks like SIDD, as mentioned in \textbf{Section} \ref{section:introduction}. Furthermore, with the increase in adaptation time, the performance deteriorates due to the continuous change of distributions. TENT-continual exhibits significantly poorer performance on average, even falling below the source DPDNet-S baseline of 5.983 dB. SAR and TENT-online* encounter similar issues due to entropy minimization. However, they employ reset strategies to prevent error accumulation, resulting in better performance compared to TENT-continual. While CauSiam achieves the best results without the resetting mechanism, demonstrating its efficacy in confronting continuously changing domains.

\begin{figure*}
\centering
\includegraphics[width=0.9\linewidth]{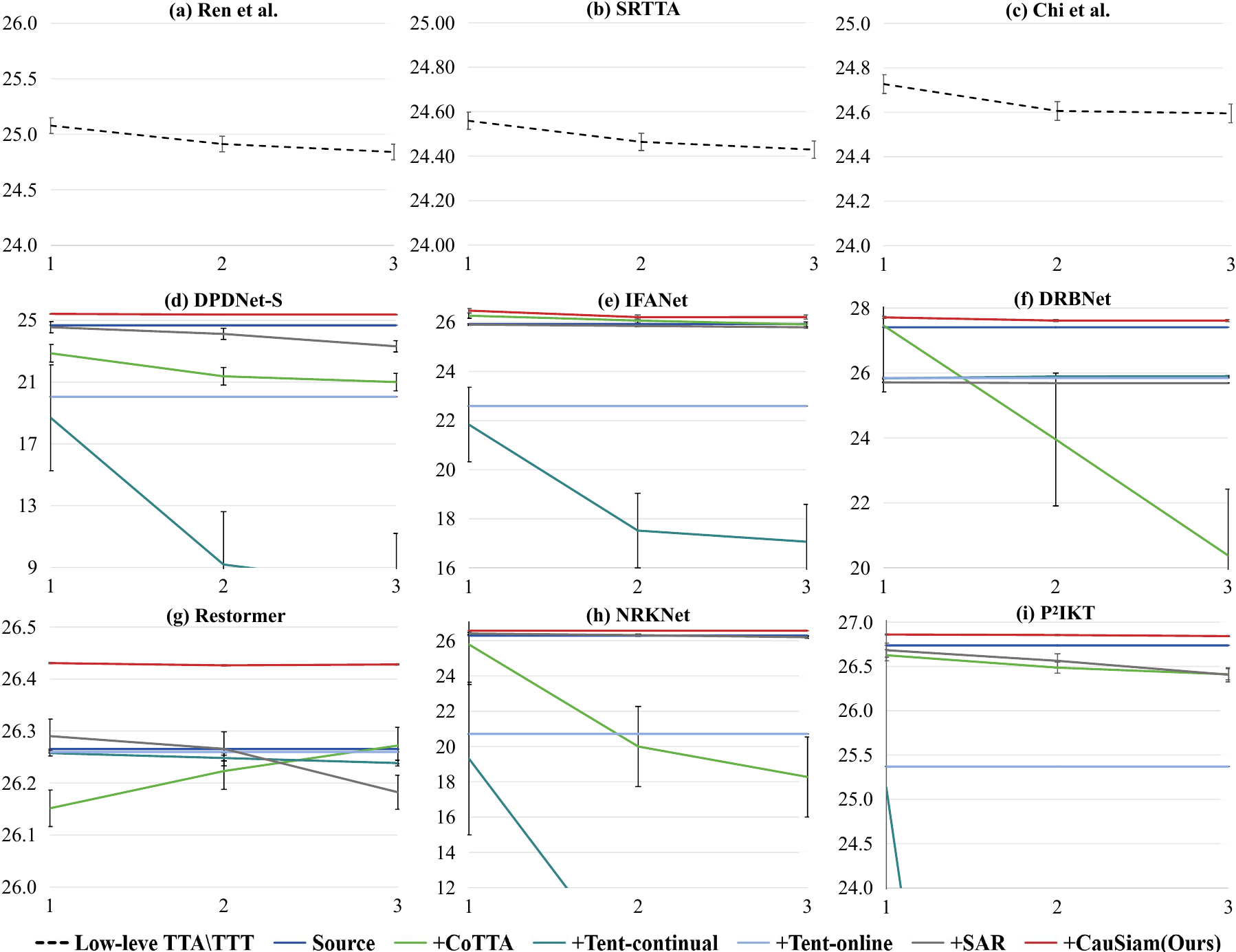}
\caption{Performance of low-level TTA$\backslash$TTT, CTTA and our CauSiam methods on different source SIDD models during long-term continual test-time adaptation. The horizontal axis represents the current round number, while the vertical axis denotes the average PSNR values of four test sets in the current round.}
\label{fig:expreiment-long-avg}
\end{figure*}

\subsection{Qualitative Comparison}
Fig. \ref{fig:qualitative-exp-sidd} shows qualitative comparisons between SIDD models with and without our CauSiam during continuous adaptation on the DPDD, RealDoF, LFDOF, RTF, and CHUK test sets. Overall, CauSiam prioritizes eliminating artifacts and generating authentic and reliable details. The analysis follows the chronological sequence of continuous adaptation. It can be observed that: 1) For DPDD, CauSiam corrects structural errors and shape distortions produced by DPDNet-S, IFANet, DRBNet, and NRKNet source models, particularly in regions with black and white stripes, and addresses the false colors generated by Restormer, such as turning white areas to light yellow; 2) In RealDOF, the NRKNet model intensifies blur on the railing, while CauSiam distinguishes its shape and produces clearer structures. Similar phenomena can also be observed in the P$^2$IKT model. It is attributed to existing methods lack of semantic priors and adaptive adjustment for lens-specific PSF heterogeneity; 3) LFDOF is captured by a light field camera and represents an out-of-distribution scenario with prominent lens-specific PSF heterogeneity, as shown in Fig. \ref{fig:movitation}(a). DRBNet effectively restores DPDD images (lens-agnostic), but it causes severe image collapse in the middle petals of LFDOF images (lens-specific). Similar issues exist in the other four SIDD methods. This further clarifies that lens-specific PSF heterogeneity is a key factor contributing to the generalization degradation of the benchmark SIDD approaches, as discussed in \textbf{Section} \ref{section:introduction}; 4) For the RTF dataset, DPDNet-S over-restores and introduces artifacts. CauSiam mitigates these artifacts, producing results closer to the ground truth; 5) In the CHUK dataset, source models like DPDNet-S, IFANet, and DRBNet introduce noise and false information, while CauSiam, with its causal-driven semantic prior integration module, eliminates false semantics and enhances genuine textures. 

Fig. \ref{fig:qualitative-exp-ctta} illustrates the qualitative comparison of the source SIDD model, four CTTA methods, and our CauSiam during continuous adaptation. Overall, incorporating CauSiam excels in preserving fidelity and reducing artifacts compared to other methods. Specifically, it is evident that: 1) CauSiam eliminates false black shadows around street lamps in DPDD, removes the white wheelchair artifacts on the blue sign in LFDOF, and reduces noise near glasses in CHUK; 2) CauSiam generates more intricate structures, such as the window frames in RealDoF and texture details on rocks in RTF. Additionally, it achieves more accurate colors compared to other CTTA algorithms like CoTTA and TENT-continual. Over time, images recovered by CoTTA and TENT-continual exhibit increasingly cyan and yellow hues respectively, indicating that they are incapable of addressing CTTA challenges like catastrophic forgetting and error accumulation. In \textbf{Section} \ref{subsection:long-ctta}, long-term CTTA experiments further validate that the performance of CoTTA and TENT-continual degrades over time, eventually becoming worse than the source model. In contrast, our algorithm consistently produces stable and visually friendly results. 

\begin{table*}
\caption{{Performance of low-level TTA, low-level TTT, CTTA, and our CauSiam methods on SIDD task during long-term continual test-time adaptation. PSNR (dB) is used as the evaluation metric. We repeat the sequence of four SIDD test datasets in three rounds. A is the average of PSNR. $\uparrow$ indicates that higher values correspond to better quality. \textcolor{myred}{+}(\textcolor{myblue}{-}) denotes the \textcolor{myred}{improvement} (\textcolor{myblue}{reduction}) in performance relative to the ``Source Only" baseline. \textbf{Bold} indicates the best performer and \underline{underline} indicates the second-best.}}
\centering
\tabcolsep=0.06cm
\resizebox{\textwidth}{!}{
\begin{tabular}{l|llll|llll|llll|l}
\toprule[1pt]
Time & \multicolumn{12}{c}{t $\xrightarrow{\hspace{12cm}}$}  \\
\midrule
Round & \multicolumn{4}{c|}{1} & \multicolumn{4}{c|}{{2}} & \multicolumn{4}{c|}{3} & \multirow{2}{*}{Average} \\
Methods & \rotatebox{60}{DPDD} & \rotatebox{60}{RealDOF} & \rotatebox{60}{LFDOF} & \rotatebox{60}{RTF} & \rotatebox{60}{DPDD} & \rotatebox{60}{RealDOF} & \rotatebox{60}{LFDOF} & \rotatebox{60}{RTF} & \rotatebox{60}{DPDD} & \rotatebox{60}{RealDOF} & \rotatebox{60}{LFDOF} & \rotatebox{60}{RTF} & PSNR$\uparrow$ \scriptsize Gains\\
\midrule[1pt]
{Ren et al.}  & {23.686} & {22.259} & {25.459} & {23.728} & {23.403} & {22.210} &	{25.301} & {23.432} & {23.296} & {22.185} & {25.235} & {23.184} & {24.943} 
 \\
{SRTTA} & {24.442} & {22.376} & {24.739} & {23.990} & {24.199} & {22.435} & {24.645} & {24.040} & {24.188} & {22.437} & {24.604} & {24.020} & {24.484} 
 \\
{Chi et al.}  & {24.166} & {22.244} & {24.993} & {23.556} & {23.884} & {22.220} & {24.876} & {23.626} & {23.769} & {22.260} & {24.872} & {23.638} &
{24.643} 
 \\
 \midrule
 {GGKMNet} & {26.039} & {24.942} & \multicolumn{1}{c}{{\textemdash}} & {25.895} & {26.039} & {24.942} & \multicolumn{1}{c}{{\textemdash}} & {25.895} & {26.039} & {24.942} & \multicolumn{1}{c}{{\textemdash}} & {25.895} & \multicolumn{1}{c}{{\textemdash}} \\
{GGKMNet$^\dag$} & {26.272} & {25.355} & \multicolumn{1}{c}{{\textemdash}} & {26.012} & {26.272} & {25.355} & \multicolumn{1}{c}{{\textemdash}} & {26.012} & {26.272} & {25.355} & \multicolumn{1}{c}{{\textemdash}} & {26.012} & \multicolumn{1}{c}{{\textemdash}}\\
\midrule
DPDNet-S & 24.648 & 23.254  &  24.810 & 23.578  &  24.648 & 23.254 & 24.810 & 23.578 & 24.648 & 23.254 & 24.810 & 23.578 & 24.676  \\
+ CoTTA & 24.633 & 22.439 & 22.746 & 21.583 & 22.224 & 20.873 & 21.332 & 20.921 & 21.660 & 20.581 & 20.977 & 20.436 & 21.745 \textcolor{myblue}{\scriptsize -2.931} \\
+ TENT-continual & 24.590 & 23.083 & 18.002 & 11.013 & 11.329 & 10.836 & 8.881 & 8.269 & 8.278 & 8.237 & 7.707 & 7.442 & 11.887 \textcolor{myblue}{\scriptsize -12.789} \\
+ TENT-online* & 24.590 & 23.240 & 19.244 & \underline{23.569} & \underline{24.590} & \underline{23.240} & 19.244 & \underline{23.569} & \underline{24.590} & \underline{23.240} & 19.244 & \underline{23.569} & 20.047 \textcolor{myblue}{\scriptsize -4.629}\\
+ SAR & \underline{24.646} & \underline{23.248} & \underline{24.663} & 23.366 & 24.485 & 23.062 & \underline{24.185} & 22.914 & 24.042 & 22.683 & \underline{23.320} & 21.999 & \underline{23.992} \textcolor{myblue}{\scriptsize -0.684}\\
\rowcolor{grayblue}
+ CauSiam(Ours) & \textbf{24.862} & \textbf{23.713} & \textbf{25.617} & \textbf{24.399} & \textbf{24.800} & \textbf{23.733} & \textbf{25.568} & \textbf{24.399} & \textbf{24.800} & \textbf{23.732} & \textbf{25.568} & \textbf{24.399} & \textbf{25.382} \textcolor{myred}{\scriptsize +0.706} \\
\midrule 
IFANet & 25.364 & 24.707 & 26.107 & 24.926 &  25.364 & 24.707 & 26.107 & 24.926  &  25.364 & 24.707 & 26.107 & 24.926  & 25.932  \\
+ CoTTA & \underline{25.704} & \underline{25.102} & \underline{26.434} & \underline{25.489} & \underline{25.354} & 24.118 & \underline{26.308} & \textbf{25.262} & 25.025 & 23.533 & \underline{26.197} & \underline{25.051} & \underline{26.086} \textcolor{myred}{\scriptsize +0.154}\\
+ TENT-continual & 25.313 & 24.672 & 21.420 & 17.210 & 19.002 & 18.111 & 17.359 & 16.365 & 17.823 & 16.978 & 17.009 & 16.538 & 18.808 \textcolor{myblue}{\scriptsize -7.124}\\
+ TENT-online* & 25.313 & 24.685 & 22.091 & 24.898 & 25.313 & 24.685 & 22.090 & 24.899 & 25.313 & 24.685 & 22.090 & 24.898 & 22.590 \textcolor{myblue}{\scriptsize -3.342}\\
+ SAR & 25.363 & 24.707 & 26.077 & 24.888 & \textbf{25.359} & \underline{24.709} & 26.015 & 24.851 & \textbf{25.354} & \underline{24.710} & 25.950 & 24.808 & 25.853 \textcolor{myblue}{\scriptsize -0.079}\\
\rowcolor{grayblue}
+ CauSiam(Ours) & \textbf{25.756} & \textbf{25.204} & \textbf{26.661} & \textbf{25.804} & 25.220 & \textbf{24.786} & \textbf{26.444} & \underline{25.223} & \underline{25.201} & \textbf{24.775} & \textbf{26.452} & \textbf{25.233} & \textbf{26.302} \textcolor{myred}{\scriptsize +0.370}\\
\midrule 
DRBNet & 25.722 & 25.743 & 27.737 & 26.221 &  25.722 & 25.743 & 27.737 & 26.221  & 25.722 & 25.743 & 27.737 & 26.221  & 27.409  \\
+ CoTTA & \textbf{25.754} & \underline{25.864} & \underline{27.795} & \underline{26.121} & \underline{25.448} & \underline{25.198} & 23.713 & 23.900 & 20.197 & 15.236 & 20.677 & 22.956 & 23.933 \textcolor{myblue}{\scriptsize -3.476} \\
+ TENT-continual & 24.241 & 22.247 & 26.258 & 25.607 & 24.113 & 22.297 & \underline{26.338} & 25.636 & 24.129 & 22.251 & \underline{26.348} & 25.630 & \underline{25.878} \textcolor{myblue}{\scriptsize -1.531}\\
+ TENT-online* & 24.241 & 23.117 & 26.206 & 25.711 & 24.241 & 23.117 & 26.206 & \underline{25.711} & \underline{24.241} & \underline{23.117} & 26.206 & \underline{25.711} & 25.846 \textcolor{myblue}{\scriptsize -1.563} \\
+ SAR & 24.237 & 22.200 & 26.117 & 25.484 & 23.965 & 22.201 & 26.118 & 25.485 & 23.966 & 22.201 & 26.118 & 25.485 & 25.698 \textcolor{myblue}{\scriptsize -1.711}\\
\rowcolor{grayblue}
+ CauSiam(Ours) & \underline{25.750} & \textbf{25.865} & \textbf{28.085} & \textbf{26.476} & \textbf{25.775} & \textbf{25.969} & \textbf{27.957} & \textbf{26.476} & \textbf{25.775} & \textbf{25.969} & \textbf{27.957} & \textbf{26.476} & \textbf{27.648} 
 \textcolor{myred}{\scriptsize +0.239} \\
\hline 
Restormer & 25.977 & 25.086 & 26.438 & 24.236 &  25.977 & 25.086 & 26.438 & 24.236  &  25.977 & 25.086 & 26.438 & 24.236  &  26.265 \\
+ CoTTA & \underline{26.039} & 24.000 & 26.353 & \textbf{24.768} & 25.780 & 24.753 & 26.416 & \textbf{24.732} & 25.810 & 24.642 & \underline{26.471} & \textbf{25.008} & 26.215 \textcolor{myblue}{\scriptsize -0.050} \\
+ TENT-continual & 25.977 & 25.086 & 26.429 & 24.230 & \underline{25.978} & 25.087 & 26.418 & 24.226 & \underline{25.979} & \underline{25.089} & 26.406 & 24.220 & 26.248 \textcolor{myblue}{\scriptsize -0.017} \\
+ TENT-online* & 25.977 & \underline{25.087} & 26.432 & 24.235 & 25.977 & 25.086 & 26.432 & 24.235 & 25.977 & 25.086 & 26.432 & 24.235 & \underline{26.260} \textcolor{myblue}{\scriptsize -0.005} \\
+ SAR & 25.978 & 25.086 & \underline{26.468} & 24.235 & 25.977 & \underline{25.096} & \underline{26.438} & 24.235 & 25.977 & 25.086 & 26.338 & 24.235 & 26.246 \textcolor{myblue}{\scriptsize -0.019} \\
\rowcolor{grayblue}
+ CauSiam(Ours) & \textbf{26.146} & \textbf{25.328} & \textbf{26.598} & \underline{24.387} & \textbf{26.145} & \textbf{25.328} & \textbf{26.593} & \underline{24.387} & \textbf{26.145} & \textbf{25.328} & \textbf{26.595} & \underline{24.388} & \textbf{26.428} \textcolor{myred}{\scriptsize +0.163} \\
\midrule
NRKNet & 26.109 & 25.027 & 26.398 & 25.949 &  26.109 & 25.027 & 26.398 & 25.949  &  26.109 & 25.027 & 26.398 & 25.949  &  26.283  \\
+ CoTTA & \underline{26.156} & 24.939 & 25.873 & 23.649 & 24.187 & 21.734 & 19.587 & 15.203 & 19.088 & 19.476 & 18.043 & 20.366 & 21.354 
\textcolor{myblue}{\scriptsize -4.929} \\
+ TENT-continual & 26.102 & 24.986 & 18.539 & 8.749 & 7.893 & 7.300 & 6.474 & 5.939 & 6.156 & 5.713 & 6.111 & 5.475 & 10.676 \textcolor{myblue}{\scriptsize -15.607} \\
+ TENT-online* & 26.102 & 25.067 & 19.690 & 25.906 & \underline{26.102} & \underline{25.067} & 19.690 & \underline{25.906} & \underline{26.102} & \underline{25.067} & 19.690 & \underline{25.906} & 20.712 \textcolor{myblue}{\scriptsize -5.571} \\
+ SAR & 26.116 & \underline{25.080} & \underline{26.543} & \underline{25.825} & 26.095 & 25.023 & \underline{26.442} & 25.737 & 26.051  & 24.942 & \underline{26.313} & 25.639 & \underline{26.304} \textcolor{myred}{\scriptsize +0.021} \\
\rowcolor{grayblue}
+ CauSiam(Ours) & \textbf{26.204} & \textbf{25.220} & \textbf{26.705} & \textbf{26.087} & \textbf{26.211} & \textbf{25.222} & \textbf{26.704} & \textbf{26.087} & \textbf{26.211} & \textbf{25.222} & \textbf{26.704} & \textbf{26.087} & \textbf{26.561} \textcolor{myred}{\scriptsize +0.278} \\
\midrule
{P$^2$IKT}&
{26.284} & {25.479} & {26.897} & {25.883} &
{26.284} & {25.479} & {26.897} & {25.883} &
{26.284} & {25.479} & {26.897} & {25.883} &
{26.737}\\
{+ CoTTA}&
\underline{{26.344}} & {25.360} & {26.754} & \underline{{25.929}} &
{26.280} & {25.245} & {26.614} & {25.796} &
{26.087} & {25.093} & {26.556} & {25.766} &
{26.505} \textcolor{myblue}{\scriptsize -0.232}\\
{+ TENT-continual}&
{26.291} & {25.331} & {25.079} & {22.460} &
{21.384} & {19.177} & {10.460} & {7.140} &
{9.324} & {8.159} & {7.573} & {6.595} &
{14.898} \textcolor{myblue}{\scriptsize -11.839}\\
{+ TENT-online*}&
{26.291} & {25.440} & {25.253} & {25.859} &
{26.291} & \underline{{25.440}} & {25.253} & \underline{{25.859}} &
\underline{{26.291}} & \underline{{25.440}} & {25.253} & \underline{{25.859}} &
{25.370} \textcolor{myblue}{\scriptsize -1.368}\\
{+ SAR}&
{26.285} & \underline{{25.472}} & \underline{{26.835}} & {25.806} &
\underline{{26.298}} & {25.371} & \underline{{26.700}} & {25.704} &
{26.265} & {25.216} & \underline{{26.528}} & {25.580} &
\underline{{26.551}} \textcolor{myblue}{\scriptsize -0.186}\\
\rowcolor{grayblue}
{+ CauSiam(Ours)}&
\textbf{{26.378}} & \textbf{{25.664}} & \textbf{{27.025}} & \textbf{{26.151}} &
\textbf{{26.378}} & \textbf{{25.565}} & \textbf{{27.021}} & \textbf{{26.111}} &
\textbf{{26.378}} & \textbf{{25.644}} & \textbf{{27.001}} & \textbf{{26.053}} &
\textbf{{26.857}} \textcolor{myred}{\scriptsize +0.120} \\
\bottomrule[1pt]
\end{tabular}}
\label{table:long-quantitative-comparison}
\end{table*}

\subsection{Comparison in the Long-Term Adaptation} 
\label{subsection:long-ctta}
To evaluate the performance of low-level TTA$\backslash$TTT, CTTA, and our CauSiam methods under the long-term continual test-time adaptation scenario \citep{wang2022continual}, we conduct experiments over three rounds with the same target domains, as shown in Table \ref{table:long-quantitative-comparison} and Fig. \ref{fig:expreiment-long-avg}. 

The last column of Table \ref{table:long-quantitative-comparison} presents the average performance across three rounds, emphasizing the model's error accumulation over multiple distinct domains. Except for CauSiam, the three-round average PSNR of other CTTA algorithms is even inferior to the source baseline. Low-level TTA and TTT methods encounter similar challenges. In contrast, after multiple rounds of testing, CauSiam achieves a maximum PSNR increase (0.706 dB) compared with the previous SOTA CoTTA method, demonstrating its ability to continuously adapt to dynamic domains in the SIDD task.

\begin{table*}
\caption{Contributions of each component in CauSiam. ``VSPI" represents the VLMs-guided semantic priors integration module. ``$L_s$" denotes the spatial consistency loss, while ``$L_f$" refers to the high-frequency consistency loss. ``$\text{CA}_\text{only}$" represents only the cross attention module updated during the back-propagation of the online model, and ``EMA" refers to the parameter update approach of the offline model. The best performers are highlighted in \textbf{bold}.}
\centering
\tabcolsep=0.1cm
\begin{tabular}{cccccc|ll|ll|ll|ll|ll}
\toprule[1pt]
&\multicolumn{5}{l|}{Time} & \multicolumn{8}{c}{t}{$\xrightarrow{\hspace{7.5cm}}$} \\
\midrule
\multicolumn{6}{c|}{Method} &\multicolumn{2}{c|}{DPDD} & \multicolumn{2}{c|}{RealDOF} & \multicolumn{2}{c|}{LFDOF} & \multicolumn{2}{c|}{RTF} & \multicolumn{2}{c}{Average} \\
& VSPI & $L_s$ & $L_f$ & $\text{CA}_\text{only}$ & EMA  & PSNR & SSIM & PSNR & SSIM & PSNR & SSIM & PSNR & SSIM & PSNR$\uparrow$ & SSIM$\uparrow$ \\
\midrule[1pt]
(a) & & & & &  & 24.648 & 0.758 & 23.254 & 0.686 & 24.810 & 0.768 & 23.578 & 0.757 & 24.676 & 0.762\\
(b) & \textcolor{gray}{\XSolidBrush} & \Checkmark & \Checkmark & \Checkmark & \Checkmark & \textbf{24.984} & \textbf{0.766} & 23.711 & 0.686 & 25.209 & 0.776 & 24.148 & 0.779 & 25.077 & 0.770 \\
(c) & \Checkmark & \textcolor{gray}{\XSolidBrush} & \Checkmark & \Checkmark & \Checkmark  & 24.390 & 0.752 & 23.274 & 0.679 & 25.114 & 0.766 & 23.729 & 0.761 & 24.910 & 0.760 \\
(d) & \Checkmark & \Checkmark & \textcolor{gray}{\XSolidBrush} & \Checkmark & \Checkmark & 24.782 & 0.755 & 23.513 & 0.682 & 25.606 & 0.775 & 24.395 & \textbf{0.785} & 25.384 & 0.768 \\
(e) &\Checkmark & \Checkmark & \Checkmark &  \textcolor{gray}{\XSolidBrush} & \Checkmark &
24.457 & 0.740 & 22.959 & 0.663 & 25.402 & 0.777 & 24.072 & 0.769 & 25.146 & 0.767 \\
(f) & \Checkmark & \Checkmark & \Checkmark & \Checkmark & \textcolor{gray}{\XSolidBrush} & 24.350 & 0.748 & 23.125 & 0.675 & 25.605 & \textbf{0.779} & 24.398 & \textbf{0.785} & 25.323 & 0.770 \\
\rowcolor{grayblue}
(g) &\Checkmark & \Checkmark & \Checkmark & \Checkmark & \Checkmark  & 24.862 & 0.759 & \textbf{23.713} & \textbf{0.687} & \textbf{25.617} & 0.778 & \textbf{24.399} & \textbf{0.785} & \textbf{25.412} & \textbf{0.771} \\
\bottomrule[1pt]
\end{tabular}
\label{table:ablation-component}
\end{table*}

\begin{figure}
\centering
\includegraphics[width=\linewidth]{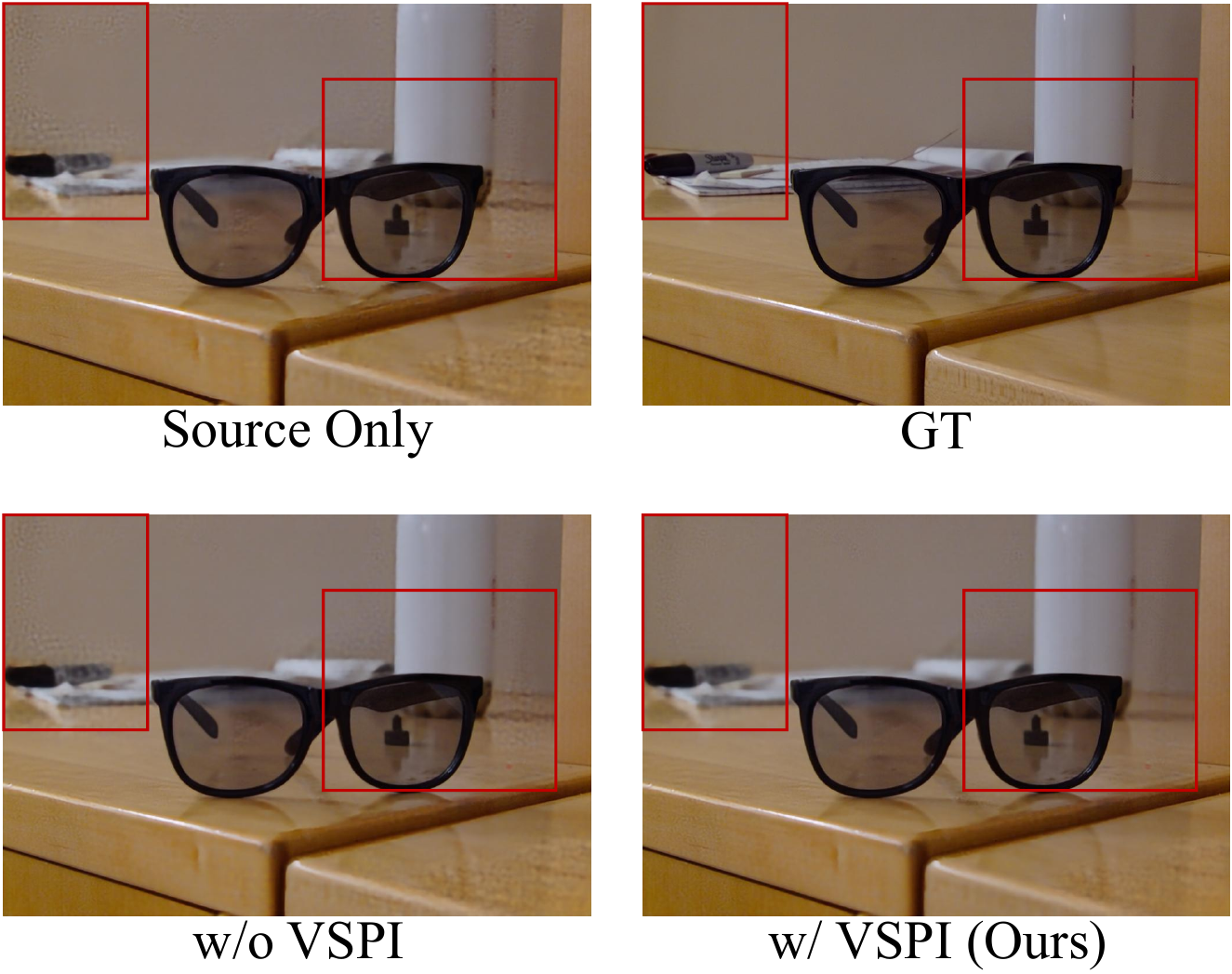}
\caption{{Effects of VLMs-guided semantic priors integration (VSPI) module.}}
\label{fig:exp-ablation-cspi}
\end{figure}

\begin{figure}
\centering
\includegraphics[width=0.75\linewidth]{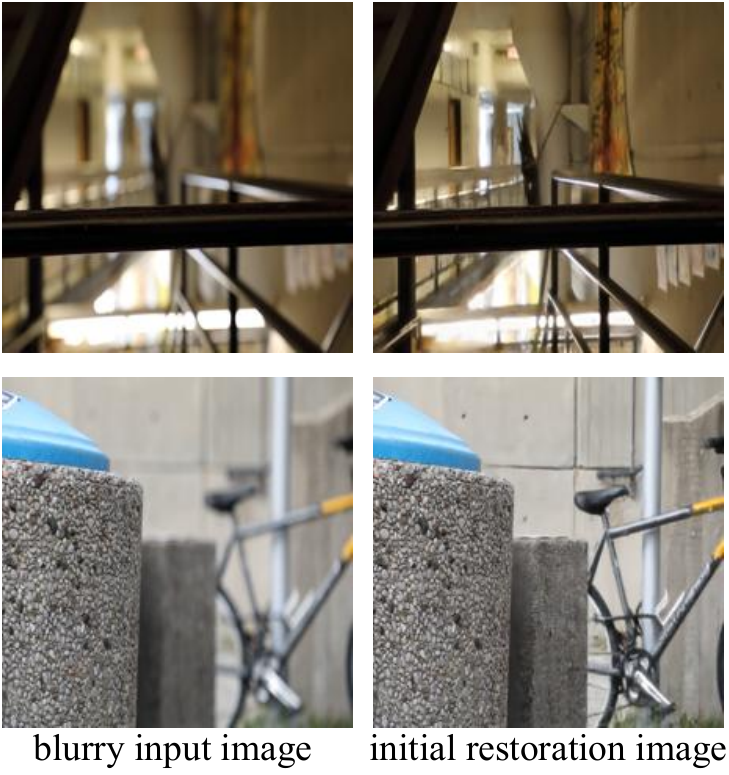}
\caption{{Visualization of different input options for the VLM module, including the blurry input image and the restored image from the source model.}}
\label{fig:without-restored-model}
\end{figure}

Fig. \ref{fig:expreiment-long-avg} visually illustrates the performance across different rounds, reflecting the extent of the model's catastrophic forgetting. {Low-level TTA or TTT algorithms ( Ren et al., SRTTA, and Chi et al.) experience performance degradation issues. TENT-continual suffers from severe performance degradation, while CoTTA and SAR decrease slightly.} Notably, TENT-online* maintains its performance by using additional domain information and resetting to the initial trained model when encountering a new domain. However, this is often not available in real-world SIDD scenarios. The results indicate that both optimizing loss functions (CoTTA) and resetting mechanism (CoTTA, SAR, TENT-online*) are effective for mitigating challenges, while TENT-continual with vanilla entropy-minimization loss is extremely unsuitable for SIDD tasks. CauSiam effectively alleviates catastrophic forgetting and error accumulation in CTTA through knowledge transfer mechanisms such as EMA and only updating CA module that integrates semantic priors, achieving minimal performance degradation.

\begin{table*}
\caption{Ablation study of semantic priors from different VLMs.}
\centering
\tabcolsep=0.1cm
\begin{tabular}{l|ll|ll|ll|ll|ll}
\toprule[1pt]
Time & \multicolumn{8}{c}{t}{$\xrightarrow{\hspace{7.5cm}}$} \\
\midrule
\multirow{2}{*}{Method}&\multicolumn{2}{c|}{DPDD} & \multicolumn{2}{c|}{RealDOF} & \multicolumn{2}{c|}{LFDOF} & \multicolumn{2}{c|}{RTF} & \multicolumn{2}{c}{Average} \\
& PSNR & SSIM & PSNR & SSIM & PSNR & SSIM & PSNR & SSIM & PSNR$\uparrow$ & SSIM$\uparrow$ \\
\midrule[1pt]
Source Only & 24.648 & 0.758 & 23.254 & 0.686 & 24.810 & 0.768 & 23.578 & 0.757 & 24.676 & 0.762\\
\rowcolor{grayblue}
+Ours-CLIP \citep{radford2021learning}  & 24.862 & 0.759 & 23.713 & 0.687 & 25.617 & 0.778 & 24.399 & 0.785 & 25.412 & 0.771 \\
+Ours-BLIP\_2 \citep{li2023blip} & 24.647 & 0.756 & 22.893 & 0.670 & 25.044 & 0.781 & 24.147 & 0.784 & 24.864 & 0.773 \\
+Ours-DINOv2 \citep{oquab2024dinov} & 24.984 & 0.766 & 23.714 & 0.700 & 25.209 & 0.779 & 24.148 & 0.779 & 25.077 & 0.774 \\
+Ours-DA-CLIP \citep{luo2024controlling} & 24.907 & 0.760 & 23.705 & 0.691 & 25.526 & 0.782 & 24.211 & 0.771 & 25.335 & 0.774 \\
+Ours-Depth \citep{yang2024depth} & 24.984 & 0.766 & 23.714 & 0.699 & 25.198 & 0.779 & 24.148 & 0.779 & 25.068 & 0.774 \\
\bottomrule[1pt]
\end{tabular}
\label{table:ablation-different-priors}
\end{table*}

\begin{table*}
\caption{Ablation study on different backbones of CLIP image encoder.}
\centering
\tabcolsep=0.1cm
\begin{tabular}{l|ll|ll|ll|ll|ll}
\toprule[1pt]
Time & \multicolumn{8}{c}{t}{$\xrightarrow{\hspace{7cm}}$} \\
\midrule
\multirow{2}{*}{Backbone}&\multicolumn{2}{c|}{DPDD} & \multicolumn{2}{c|}{RealDOF} & \multicolumn{2}{c|}{LFDOF} & \multicolumn{2}{c|}{RTF} & \multicolumn{2}{c}{Average} \\
& PSNR & SSIM & PSNR & SSIM & PSNR & SSIM & PSNR & SSIM & PSNR$\uparrow$ & SSIM$\uparrow$ \\
\midrule
 ResNet-50 & 24.990 & 0.765 & 23.721 & 0.698 & 25.213 & 0.778 & 24.083 & 0.776 & 25.079 & 0.773 \\
 ResNet-101 & 24.965 & 0.765 & 23.670 & 0.697 & 25.249 & 0.780 & 24.181 & 0.781 & 25.107 & 0.774 \\
 ViT-B/16 & 24.899 & 0.764 & 23.574 & 0.695 & 25.006 & 0.775 & 24.114 & 0.790 & 24.892 & 0.770 \\
\rowcolor{grayblue}
 ViT-B/32 & 24.862 & 0.759 & 23.713 & 0.687 & 25.617 & 0.778 & 24.399 & 0.785 & 25.412 & 0.771 \\
\bottomrule[1pt]
\end{tabular}
\label{table:ablation-backbone-clip}
\end{table*}

\subsection{Ablation Study}
\label{subsection:experiments-ablation}
To assess the contribution of each component and hyper-parameter in enhancing the generalization of SIDD during CTTA, we conduct comprehensive ablation studies. We use DPDNet-S \citep{abuolaim2020defocus}, trained on the DPDD training set, as the baseline SIDD model, referred to as ``\textbf{Source Only}".

\textbf{Impact of the VLMs-guided semantic priors integration module (VSPI).} 
As depicted in Table \ref{table:ablation-component}(b) and (g), integrating VLMs-guided semantic priors improves the average PSNR by 0.335 dB. Fig. \ref{fig:exp-ablation-cspi} demonstrates that VSPI module effectively eliminates semantically erroneous textures introduced by the source model, such as artifacts on the left wall and the right table. These results underline the effectiveness of VSPI in our proposed CauSiam, which can be attributed to ensuring causal identifiability between the blurry inputs and restored images. Furthermore, we conduct ablation experiments to validate our design choice to use the initial restoration $\mathcal{F}_{\zeta}(x_{\text{test}})$ from the source model as the input for the VLMs. As illustrated in Fig. \ref{fig:without-restored-model}, employing the initial restoration image as input to VLMs effectively mitigates the effects of blur degradation. This design enables the VLM to leverage enhanced details and textures from the restoration process while integrating semantic information for further refinement.

\textbf{Impact of different loss components.} 
As shown in Table \ref{table:ablation-component}(c) and (d), we evaluate individual impacts of $\mathcal{L}_s$ and $\mathcal{L}_f$ in Equation \eqref{equ:loss_total}. The results indicate that spatial consistency loss $\mathcal{L}_s$ significantly improves model performance, with high-frequency consistency loss $\mathcal{L}_f$ providing complementary enhancements. Overall, consistency loss is essential for CTTA tasks, ensuring the network's robustness against minor input perturbations. This supports the assumption that a well-generalized model should have decision boundaries in low-density regions \cite{chapelle2005semi}.

\textbf{Impact of the knowledge transfer mechanism.} As shown in Table \ref{table:ablation-component}(e) and (f), the knowledge transfer mechanism contributes to a more stable update of CauSiam during long-term CTTA. For instance, only updating CA module improves the average PSNR by 0.266 dB compared to updating all parameters of the online model. Furthermore, employing EMA to update the offline model further enhances performance.

\textbf{Comparison of semantic priors from different VLMs.} Table \ref{table:ablation-different-priors} compares SIDD results of our CauSiam model using different VLMs to derive universal semantic priors. VLMs include CLIP \citep{radford2021learning}, BLIP\_2 \citep{li2023blip}, DINOv2 \citep{oquab2024dinov}, DA-CLIP \citep{luo2024controlling}, and Depth Anything \citep{yang2024depth}. The performance of CauSiam is influenced by VLMs, yet it consistently outperforms the benchmark ``Source Only" model across various VLMs, demonstrating its robustness. According to the causal analysis of the SCM in Fig. \ref{fig:scm}, given an input image $X$, the VLM estimates the unobservable semantic-level knowledge $K_S$ by $S$. This enables the causal effect between $X$ and the ground truth $Y$ identifiable, leading to performance improvement. Notably, BLIP\_2 and DINOv2 have complex architectures and numerous parameters, resulting in slower forward processing than CLIP. Considering both effectiveness and efficiency, we choose CLIP image encoder for extracting semantic priors.

\begin{table}
\caption{Ablation study of hyper-parameters $\alpha$, $\eta$ and $lr$, using PSNR (dB) as the evaluation metric.}
\centering
\tabcolsep=0.1cm
\begin{tabular}{cc|cccc|c}
\toprule[1pt]
\multicolumn{2}{l|}{Time} & \multicolumn{4}{c}{t}{$\xrightarrow{\hspace{3.8cm}}$} \\
\midrule
\multicolumn{2}{c|}{Parameter} & \rotatebox{30}{DPDD} & \rotatebox{30}{RealDOF} & \rotatebox{30}{LFDOF} & \rotatebox{30}{RTF} & Average \\
\midrule
$\alpha$ & 0.01 & 24.997 & 23.772 & 25.312 & 24.253 & 25.170 
\\
\rowcolor{grayblue}
$\alpha$ & 0.05 & 24.862 & 23.713 & 25.617 & 24.399 & 25.412 \\
$\alpha$ & 0.1 & 24.302 & 23.167 & 25.512 & 23.931 & 25.233 \\
$\alpha$ & 0.15 & 23.324 & 22.141 & 24.792 & 23.024 & 24.467 \\
$\alpha$ & 0.2 & 22.224 & 20.943 & 23.842 & 22.056 
& 23.490 \\
\midrule
\rowcolor{grayblue}
$\eta$ & 0.9 & 24.862 & 23.713 & 25.617 & 24.399 & 25.412 \\
$\eta$ & 0.8 & 24.828 & 23.649 & 25.622 & 24.399 & 25.409 \\
$\eta$ & 0.7 & 24.785 & 23.591 & 25.624 & 24.399 & 25.404 \\
$\eta$ & 0.6 & 24.734 & 23.531 & 25.625 & 24.399 & 25.397 \\
$\eta$ & 0.5 & 24.675 & 23.467 & 25.624 & 24.399 & 25.387 \\
\midrule
$lr$ & 5e-5 & 24.874 & 23.758 & 25.606 & 24.399 & 25.406 \\
\rowcolor{grayblue}
$lr$ & 1e-4 & 24.862 & 23.713 & 25.617 & 24.399 & 25.412 \\
$lr$ & 5e-4 & 24.801 & 23.498 & 25.649 & 24.399 & 25.421 \\
$lr$ & 1e-3 & 24.717 & 23.223 & 25.672 & 24.400 & 25.417 \\
$lr$ & 5e-3 & 23.050 & 21.171 & 25.189 & 24.405 & 24.753 \\
\bottomrule[1pt]
\end{tabular}
\label{table-ablation-hyperparameter}
\end{table}

\begin{table}
\caption{Ablation study on the iteration number K of CauSiam, using PSNR (dB) as the evaluation metric.}
\centering
\tabcolsep=0.1cm
\begin{tabular}{c|cccc|c}
\toprule[1pt]
\multicolumn{1}{c|}{Time} & \multicolumn{4}{c}{t}{$\xrightarrow{\hspace{3.8cm}}$} \\
\midrule
 K & \rotatebox{30}{DPDD} & \rotatebox{30}{RealDOF} & \rotatebox{30}{LFDOF} & \rotatebox{30}{RTF} & Average \\
\midrule
\rowcolor{grayblue}
 1 & 24.862 & 23.713 & 25.617 & 24.399 & 25.412\\
 2 & 24.886 & 23.774 & 25.577& 24.399& 25.384 \\
 3 & 24.895 & 23.758 & 25.570 & 24.399 &  25.378 \\
 4 & 24.884 & 23.744 & 25.569 & 24.399 & 25.375\\
\bottomrule[1pt]
\end{tabular}
\label{table:ablation-iteration}
\end{table}

\textbf{Different backbones of CLIP image encoder.} Table \ref{table:ablation-backbone-clip} presents results of ablation study on four different CLIP visual backbones: ResNet-50 \citep{he2016deep}, ResNet-101 \citep{he2016deep}, ViT-B/16 \citep{dosovitskiy2020image}, and ViT-B/32 \citep{dosovitskiy2020image}. Our CauSiam, with all these backbones, consistently achieves significant improvements. Notably, CLIP with ViT-B/32 backbone achieves the best average PSNR gains. Hence, we select ViT-B/32 as the backbone for CLIP image encoder.

\textbf{Analysis of hyper-parameters $\alpha$, $\eta$ and $lr$.} We conduct experiments to analyze hyper-parameters, as presented in Table \ref{table-ablation-hyperparameter}. The weight of semantic-aware features $\alpha$ serves as a bridge between the trainable CA module and the other frozen network, making its selection crucial. The best results are obtained when $\alpha$ is set to 0.05. We provide additional guidelines for choosing this parameter when adapting to new datasets. Intuitively, a relatively large value of $\alpha$ (e.g., $0.05 \leq \alpha \leq 0.1$) indicates greater incorporation of semantic knowledge from VLMs, thereby yielding superior performance on datasets with severe distribution shifts. In contrast, a relatively small $\alpha$ (e.g., $0 \leq \alpha \leq 0.05$) generally performs better on datasets with minimal distribution shifts or identical distribution. As the decay rate of EMA ($\eta$) increases, the average PSNR gradually declines, leading us to select $\eta = 0.9$ for optimal performance. The learning rate ($lr$) shows minimal impact on SIDD results when set below 1e-3, with all configurations demonstrating substantial improvements. Balancing performance across four SIDD datasets, we choose $lr$ = 1e-4 as the best configuration. Although $lr$ = 5e-4 yields the highest average PSNR, its performance on RealDOF is limited.

\textbf{The iteration number $K$ for CauSiam.} We investigate the effect of the iteration number $K$ for CauSiam during CTTA. As depicted in Table \ref{table:ablation-iteration}, the performance across four SIDD datasets remains stable for iterations 1, 2, 3, and 4. Since optimal performance is achieved with one iteration, we set the iteration number to one for maximum efficiency.

\begin{table}
\caption{{The detailed analysis of computational complexity.}}
\centering
\tabcolsep=0.06cm
\begin{tabular}{lcccc}
\toprule[1pt]
\multirow{2}{*}{{Method}} & {Run Time} & {GPU Memory} & {FLOPs} \\
& {(seconds/image)} & {(GB)}  & {(G)} \\
\bottomrule[1pt]
 {Ren et al.} & {1.636} & {10.705} & {54.719}\\
 {SRTTA} & {9.332} & {14.072} & {329.017} \\
 {Chi et al.} & {5.553} & {21.680} & {131.988} \\
 \midrule
 {GGKMNet} & {{\textemdash}} & {{\textemdash}} & {{\textemdash}}\\
 {GGKMNet$^\dag$} & {{\textemdash}} & {{\textemdash}} & {{\textemdash}}\\
 \midrule
 {DPDNet-S} & {0.151} & {6.834} & {54.719} \\
 {+ TENT-continua} & {0.273} & {8.922} & {54.728} \\
 {+ TENT-online*} & {0.273} & {8.922} & {54.728} \\
 {+ SAR} & {0.421} & {9.086} & {109.456} \\
 {+ CoTTA} & {2.193} & {13.842} & {328.317}\\
 {+ CauSiam(Ours)} & {2.581} & {12.627} & {389.207} \\
 \midrule
  {IFANet} & {0.011} & {3.938} & {29.733}\\
  {+ TENT-continual} & {0.107} & {9.771} & {29.734}\\
  {+ TENT-online*}  & {0.107} & {9.771} & {29.734} \\
  {+ SAR}  & {0.369} & {9.852} & {59.468} \\
  {+ CoTTA}  & {1.449} & {12.314} & {178.402} \\
  {+ CauSiam(Ours)} & {1.652} & {11.723} & {224.032} \\
  \midrule
  {DRBNet} & {0.008} & {8.740} & {49.158}\\
  {+ TENT-continual} & {0.023} & {12.338} & {49.159}\\
  {+ TENT-online*} & {0.023} & {12.338} & {49.159} \\
  {+ SAR} & {0.419} & {12.383} & {98.319}\\
  {+ CoTTA} & {2.035} & {13.439} & {294.950} \\
  {+ CauSiam(Ours)} & {2.748} & {12.254} & {357.509} \\
  \midrule
  {Restormer} & {0.808} & {22.449} & {140.990} \\
  {+ TENT-continual} & {1.090} & {22.449} & {140.991} \\
  {+ TENT-online*} & {1.090} & {22.449} & {140.991} \\
  {+ SAR} & {1.497} & {22.649} & {281.984} \\
  {+ CoTTA} & {12.308} & {22.960} & {845.941} \\
  {+ CauSiam(Ours)} & {14.141} & {23.569} & {908.553} \\
  \midrule
  {NRKNet} & {0.023} & {7.537} & {78.623}\\
  {+ TENT-continual} & {0.218} & {12.143} & {78.645} \\
  {+ TENT-online*} & {0.218} & {12.143} & {78.645} \\
  {+ SAR} & {0.406} & {12.168} & {157.290}  \\
  {+ CoTTA} & {2.747} & {19.500} & {471.737}\\
  {+ CauSiam(Ours)} & {1.569} & {10.723} & {562.552} \\
  \midrule
  {P$^2$IKT} & {0.039} & {6.324} & {84.602}\\
  {+ TENT-continual} & {0.330} & {8.564} & {84.624} \\
  {+ TENT-online*} & {0.330} & {8.564} & {84.624} \\
  {+ SAR} & {0.617} & {8.575} & {169.248} \\
  {+ CoTTA} & {4.426} & {20.520} & {507.612} \\
  {+ CauSiam(Ours)} & {4.092} & {14.971} & {607.301}\\
\bottomrule[1pt]
\end{tabular}
\label{table:computational_cost}
\end{table}

\subsection{Computational Complexity}

To assess the computational efficiency, we conduct a comprehensive analysis of computational complexity across all compared baselines. This analysis includes detailed measurements of wall-clock run time, GPU memory consumption, and floating point operations per second (FLOPs). We evaluate wall-clock run time and GPU memory usage on the DPDD test set, which has a resolution of \(3 \times 1680 \times 1120\). We compute FLOPs based on input dimensions of (3,256,256) on a single NVIDIA RTX 3090 GPU. The results, summarized in Table \ref{table:computational_cost}, provide a direct comparison of our method with baselines, including low-level TTA and TTT methods \citep{ren2020video,deng2023efficient,chi2021test}, SIDD methods \citep{abuolaim2020defocus,lee2021iterative,ruan2022learning,zamir2022restormer,quan2023neumann,tang2024prior}, and high-level CTTA methods \citep{wang2021tent,niu2023towards,wang2022continual}.

Our method is comparable or even faster efficiency compared to low-level TTA methods such as Ren et al. and SRTTA, as CauSiam does not require multiple iterations at test time. In contrast, the standard setting for SRTTA involves ten iterations for convergence, significantly increasing run time. When compared to other CTTA methods, such as TENT and SAR, our approach has relatively slower computational efficiency. However, it achieves substantially better performance. CoTTA utilizes multiple augmentations and exhibits similar computational efficiency to Causiam. The key difference is that CoTTA updates all model parameters during inference, while our approach limits updates to CA module, which enhances both stability and efficiency.

Compared to source SIDD models, the incorporation of VLMs and multiple augmentations could raise additional computational complexity. To address this, we implement several enhancements to achieve an optimal trade-off between computational cost and restoration performance, such as the VLMs being used solely for inference and resizing images to 224 $\times$ 224.

\section{Conclusion}
\label{section:conclusion}
We conduct multiple motivation experiments, demonstrating that the intrinsic reason behind the performance degradation of the SIDD model under out-of-distribution scenarios is the lens-specific PSF heterogeneity. Thus, our work is the pioneer in establishing a Siamese networks-based continual test-time adaptation framework for the pixel-level regression task SIDD. Additionally, to further improve performance on severely degraded images and avoid introducing erroneous textures, we revisit the SIDD learning paradigm from a causality perspective and propose a novel SCM. Guided by the SCM, we propose CauSiam to incorporate the universal semantic priors into Siamese networks, thereby efficiently performing CTTA for SIDD. Moreover, experiments demonstrate the effectiveness and generalization of the proposed CauSiam.

\textbf{Limitations and future work.} Compared to other low-level TTA methods, CauSiam demonstrates comparable or even faster efficiency. However, it utilizes multiple augmentations and VLMs for inference, which increases computational complexity. Thus, exploring effective strategies to reduce the space and time complexity during inference is promising research. Furthermore, we will expand to a broader range of image restoration.

\begin{acknowledgements}
This work is supported by the Postdoctoral Fellowship Program, Grant No. GZC20232812, the China Postdoctoral Science Foundation, Grant No. 2024M753356, National Natural Science Foundation of China No. 62406313, Guangzhou-HKUST(GZ) Joint Funding Program, Grant No. 2023A03J0008, the Education Bureau of Guangzhou Municipality, 2023 Special Research Assistant Grant Project of the Chinese Academy of Sciences.
\end{acknowledgements}

\small{\noindent \textbf{Data Availability}
The DPDD dataset (\citep{abuolaim2020defocus}) is available at \url{https://drive.google.com/file/d/1Mq7WtYMo9mRsJ6I6ccXdY1JJQvwBuMuQ/view}.
The LFDOF dataset (\citep{ruan2021aifnet}) is available at \url{https://sweb.cityu.edu.hk/miullam/AIFNET/dataset/LFDOF.zip}. The RTF dataset (\citep{d2016non}) is available at \url{https://drive.google.com/file/d/1Zf48pu1k_kUZZTscpgrasBgH2xVty3Vr/view?usp=sharing}, The RealDOF dataset (\citep{lee2021iterative}) is available at \url{https://drive.google.com/file/d/1MyizebyGPzK-VeV1pKVf7OTDl_3GmkdQ/view}. The CUHK dataset (\citep{shi2014discriminative}) is available at \url{https://drive.google.com/file/d/1Mol1GV-1NNoSX-BCRTE09Sins8LMVRyl/view}.}

\bibliographystyle{spbasic} 
\bibliography{egbib}

\clearpage

\section{Appendix}
\label{sec:appendix}
\subsection{Definition}
We first give definitions to the path, the d-separation, and the backdoor criterion. From \cite{pearl2009causality}, we can obtain that: 
\begin{definition}
\label{def:1}
\textbf{Path}. A path consists of three components including the Chain Structure: $A \to B \to C$ or $A \leftarrow B \leftarrow C$, the Bifurcate Structure: $A \leftarrow B \to C$, and the Collisions Structure: $A \to B \leftarrow C$.
\end{definition}

\begin{definition}
\label{def:2}
\textbf{$d$-separation}. A path $p$ is blocked by a set of nodes $Z$ if and only if:
\begin{itemize}
\item $p$ contains a chain of nodes $A \to B \to C$ or a fork $A \leftarrow B \leftarrow C$ such that the middle node $B$ is in $Z$ (i.e., $B$ is conditioned on), or

\item $p$ contains a collider $A \to B \leftarrow C$ such that the collision node $B$ is not in $Z$, and no descendant of $B$ is in $Z$. If $Z$ blocks every path between two nodes $X$ and $Y$,  then $X$ and $Y$ are $d$-separated, conditional on $Z$, and thus are independent conditional on $Z$.
\end{itemize}
\end{definition}

\begin{definition}
\label{def:3}
\textbf{The Backdoor Criterion}. Given an ordered pair of variables $\left( {X,Y} \right)$ in a directed acyclic graph $G$, a set of variables $Z$ satisfies the backdoor criterion relative to $\left( {X,Y} \right)$ if no node in $Z$ is a descendant of $X$, and $Z$ blocks every path between $X$ and $Y$ that contains an arrow into $X$. If a set of variables of $Z$ satisfies the backdoor criterion for $X$ and $Y$, then the causal effect of $X$ on $Y$ is given by the formula:
\begin{equation}
    \begin{array}{l}
P\left( {Y = y\left| {do\left( {X = x} \right)} \right.} \right)\\
 = \sum\limits_z {P\left( {Y = y\left| {X = x,Z = z} \right.} \right)} P\left( {Z = z} \right)
\end{array}
\end{equation}
\end{definition}

\subsection{Theorem}
\begin{theorem}[Rules of $do$ Calculus \cite{pearl1995causal}] 
\label{theorem:1}
Let $G$ be a directed acyclic graph (DAG) associated with a causal model, and let $P(\cdot)$ stand for the probability distribution induced by that model. For any disjoint subsets of variables $X, Y, Z$, and $W$, we have the following rules.
\begin{itemize}
    \item \textbf{Rule 1} (Insertion/deletion of observations):
    \begin{equation}
    P(y \mid \hat{x}, z, w) = P(y \mid \hat{x}, w)\text{, if } (Y \Perp Z \mid X, W)_{G_{\overline{X}}}.
    \label{eq:rule1}
    \end{equation}

    \item \textbf{Rule 2} (Action/observation exchange):
    \begin{equation}
    P(y \mid \hat{x}, \hat{z}, w) = P(y \mid \hat{x}, z, w)\text{, if } (Y \Perp Z \mid X, W)_{G_{\overline{X} \underline{Z}}}.
    \label{eq:rule2}
    \end{equation}

    \item \textbf{Rule 3} (Insertion/deletion of actions):
    \begin{equation}
    P(y \mid \hat{x}, \hat{z}, w) = P(y \mid \hat{x}, w)\text{, if } (Y \Perp Z \mid X, W)_{G_{\overline{X}, \overline{Z(W)}}},
    \label{eq:rule3}
    \end{equation}
    where $Z(W)$ is the set of $Z$-nodes that are not ancestors of any W-node in $G_{\overline{X}}$.
\end{itemize}
\end{theorem}

\subsection{Multi-world Symbolic Derivation of Causal Effect Indentifiability} \label{app:symbolic-scm}

\begin{figure*}
	\centering  \includegraphics[width=\textwidth]{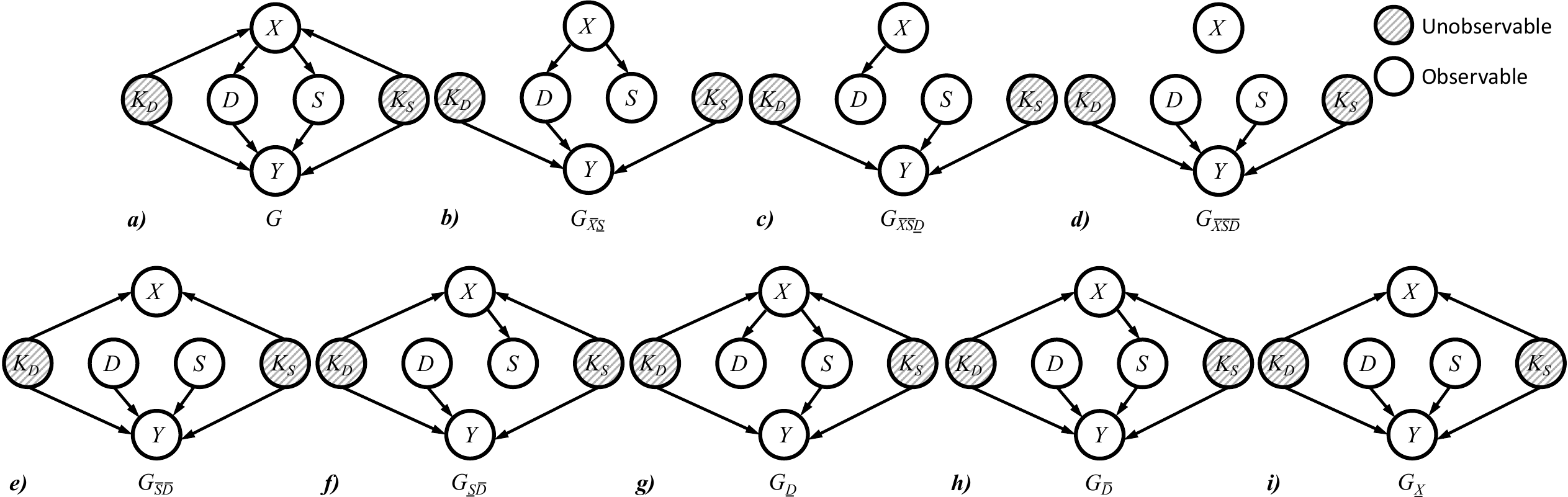}
	\caption{Subgraphs of $G$ used in the derivation of causal effects.}
	\label{fig:scm2}
	\vspace{-0.2cm}
\end{figure*}

{We compute $P(y \mid \hat{x})$ using Theorem \ref{theorem:1}. Fig. \ref{fig:scm2} illustrates the subgraphs required for the multi-world symbolic derivation.}

\noindent $P(y \mid \hat{x})$ can be expressed as: 
\begin{subequations} \label{eq:multi_world}
\begin{align}
P(y \mid \hat{x}) &= \sum_s P(y \mid \hat{x}, s) P(s \mid \hat{x}) \quad \text{(Probability)}\\
&= \sum_s \sum_d \underbrace{P(y \mid \hat{x}, s, d)}_{\textbf{1}} \underbrace{P(d \mid \hat{x}, s)}_{\textbf{2}} \underbrace{P(s \mid \hat{x})}_{\textbf{3}}. 
\end{align}
\end{subequations}

\noindent We now deal with three expressions involving $\hat{x}$, $P(y \mid \hat{x}, s, d)$, $P(d \mid \hat{x}, s)$, and $P(s \mid \hat{x})$.

\noindent The first part $P(y \mid \hat{x}, s, d)$ of Equation (\ref{eq:multi_world}b):
\begin{subequations} \label{eq:app_part1}
\begin{alignat}{2}
    P(y \mid \hat{x}, s, d) &= P(y \mid \hat{x}, \hat{s}, d) \quad \text{(Rule 2)} \quad G_{\overline{X} \underline{S}} \\
    &= P(y \mid \hat{x}, \hat{s}, \hat{d})  \quad \text{(Rule 2)} \quad G_{\overline{X} \overline{S} \underline{D}} \\
    &= P(y \mid \hat{s}, \hat{d}) \quad \quad \text{(Rule 3)} \quad G_{\overline{X} \overline{S} \overline{D}} \\
    &= \sum_{x} \underbrace{P(y \mid \hat{s}, \hat{d}, x)}_{\textbf{1.1}} \underbrace{P(x \mid \hat{s}, \hat{d})}_{\textbf{1.2}} \quad  \text{(Probability),}
\end{alignat}
\end{subequations}
{Equation (\ref{eq:app_part1}a) holds due to the conditional independence $(Y \Perp S \mid X, D)$ in $G_{\overline{X} \underline{S}}$ (Rule 2  in Theorem \ref{theorem:1}); Equation (\ref{eq:app_part1}b) holds due to $(Y \Perp D \mid X, S)$ in $G_{\overline{X} \overline{S} \underline{D}}$ (Rule 2); Equation (\ref{eq:app_part1}c) holds due to $(Y \Perp X \mid S, D)$ in $G_{\overline{X} \overline{S} \overline{D}}$ (Rule 3). We deal with two expressions in Equation (\ref{eq:app_part1}d) involving $\hat{s}$ and $\hat{d}$, $P(y \mid \hat{s}, \hat{d}, x)$ and $P(x \mid \hat{s}, \hat{d})$.}

\begin{subequations} \label{eq:app_part1_1}
\begin{align}
    P(y \mid \hat{s}, \hat{d}, x) &= P(y \mid s, \hat{d}, x) \quad \text{(Rule 2)} \quad G_{\underline{S} \overline{D}} \\
    &= P(y \mid s, d, x) \quad \text{(Rule 2)} \quad G_{\underline{D}},
\end{align}
\end{subequations}
Equation (\ref{eq:app_part1_1}a) holds due to $(Y \Perp S \mid X, D)$ in $G_{\underline{S} \overline{D}}$ (Rule 2); Equation (\ref{eq:app_part1_1}b) holds due to $(Y \Perp D \mid X, S)$ in $G_{\underline{D}}$ (Rule 2).

\begin{subequations} \label{eq:app_part1_2}
\begin{align}
    P(x \mid \hat{s}, \hat{d}) &= P(x \mid \hat{d})  \quad \text{(Rule 3)} \quad G_{\overline{S} \overline{D}} \\
    &= P(x)  \quad \quad \text{(Rule 3)} \quad G_{\overline{D}},
\end{align}
\end{subequations}
Equation (\ref{eq:app_part1_2}a) holds due to $(X \Perp S \mid D)$ in $G_{\overline{S} \overline{D}}$ (Rule 3); Equation (\ref{eq:app_part1_2}b) holds due to $(X \Perp D)$ in $G_{\overline{D}}$ (Rule 3).
Substituting Equation (\ref{eq:app_part1_1}b) and Equation (\ref{eq:app_part1_2}b) back into Equation (\ref{eq:app_part1}d) finally yields,
\begin{subequations} \label{eq:app_part1_3}
\begin{align}
    P(y \mid \hat{x}, s, d) &= \sum_{x} P(y \mid \hat{s}, \hat{d}, x) P(x \mid \hat{s}, \hat{d}) \quad  \text{(Probability)}\\
    &= \sum_{x} P(y \mid s, d, x) P(x).
\end{align}
\end{subequations}

\noindent The second part $P(d \mid \hat{x}, s)$ of Equation (\ref{eq:multi_world}b):
\begin{subequations} \label{eq:app_part2}
\begin{align}
    P(d \mid \hat{x}, s) &= P(d \mid x, s) & \text{(Rule 2)} \quad G_{\underline{X}} \\
    &= P(d \mid x)  & \text{(Independent)},
\end{align}
\end{subequations}
Equation (\ref{eq:app_part2}a) holds due to $(D \Perp X \mid S)$ in $G_{\underline{X}}$ (Rule 2), Equation (\ref{eq:app_part2}b) holds due to conditional independence.

\noindent The third part $P(s \mid \hat{x})$ of Equation (\ref{eq:multi_world}b):
\begin{subequations} \label{eq:app_part3}
\begin{align}
    P(s \mid \hat{x}) &= P(s \mid x) &\quad & \text{(Rule 2)} \quad G_{\underline{X}}
\end{align}
\end{subequations}

\noindent We take Equation (\ref{eq:app_part1_3}b), (\ref{eq:app_part2}b), and (\ref{eq:app_part3}a) back into Equation (\ref{eq:multi_world}b) obtains,
\begin{subequations} \label{eq:app_total}
\begin{align}
    P(y \mid \hat{x}) &= \sum_{s}\sum_{d} P(y \mid \hat{x},s,d) P(d \mid \hat{x},s) P(s \mid \hat{x}) \\
    &= \sum_{s}\sum_{d} \sum_{x'} P(y \mid s,d,x') P(x') P(d \mid x) P(s\mid x),
\end{align}
\end{subequations}
which is identical to Equation (\ref{eq:front_door}g) in the main paper.

\end{document}